Agence Universitaire de la Francophonie

Bureau Asie-Pacifique

**Numérique éducatif et services à la communauté universitaire francophone**

**Quadriennal 2014-2017**

# Formation à distance et outils numériques pour l'enseignement supérieur et la recherche en Asie-Pacifique (Cambodge, Laos, Vietnam)

Rapport d'expertise au compte du Bureau Asie Pacifique (BAP) de l'Agence universitaire de la Francophonie

**Partie 2**

## RECOMMANDATIONS & FEUILLE DE ROUTE

**Mokhtar BEN HENDA**

**Juin 2016**

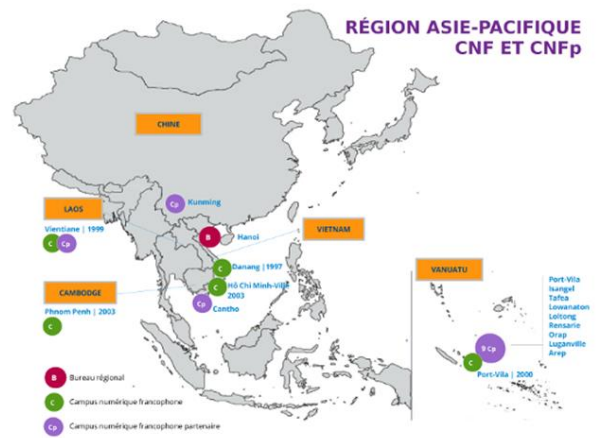

Ce rapport est le résultat d'un travail d'équipe auquel ont participé des experts nationaux et internationaux, des Chefs de Bureaux et des responsables de Campus Numériques Francophones et Campus Numériques Francophones partenaires. Chacun, de par sa position et son rôle, a contribué à la collecte de ressources d'information et à la traduction de certaines d'entre-elles, à la traduction des séances d'entretiens réalisées par l'expert international au Cambodge, Laos et Vietnam. Un remerciement leur est rendu ici comme témoignage de reconnaissance pour leur efficacité, amabilité et sens de la coopération (liste alphabétique de noms) :

Des remerciements particuliers sont adressés à Mme Sophie GOEDEFROIT, Directrice du Bureau Asie-Pacifique de l'AUF pour avoir généreusement prodigué aux membres de l'équipe ses précieux conseils et directives en vue d'atteindre les objectifs de cette expertise pour le renforcement du numérique éducatif francophone dans la région.

Ce rapport d'expertise est réalisé pour le compte du Bureau Asie Pacifique (BAP) de l'Agence universitaire de la Francophonie. Il s'inscrit dans le cadre de la mission « Numérique éducatif et services à la communauté universitaire francophone ».
V. 2.0
© BAP/AUF juin 2016

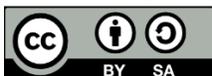

# SOMMAIRE

## ■ RAPPEL

Rappelant les termes de référence de la mission d'expertise donnant lieu à ce rapport, celle-ci se situe dans l'axe n°2 du projet « numérique éducatif » du BAP à savoir « *Appuyer le développement de projets innovants pour l'enseignement supérieur et la recherche dans les établissements membres de la région Asie-Pacifique* ».

Cette mission d'expertise a été programmée pour se dérouler en trois étapes :

1-Une revue de littérature sur les pratiques locales et recommandations de rencontres/visites pour les experts internationaux ;

2-Une mission d'expertise donnant lieu à un état des lieux, une évaluation des besoins et des recommandations ;

3-Un atelier de co-construction avec les équipes de l'AUF en Asie (BAP et Campus numériques francophones) pour définir une feuille de route 2016-2017.

**AVERTISSEMENT**

Ce rapport est en deux parties :

- La première partie est un état des lieux détaillé dans lequel le contexte éducatif du Cambodge, Laos et Vietnam ainsi que le contenu des entretiens et des visites réalisés dans 21 institutions à caractère universitaire sont rendus avec plus de détails.

- La deuxième (composée de ce document) est un ensemble de recommandations et une feuille de route proposée pour la période 2016-2017

# ◼ AVANT-PROPOS

Cette deuxième partie du rapport se ressource essentiellement dans les résultats de l'état des lieux décrit dans la première partie. Elle se fonde aussi sur un constat « préoccupant » énoncé dans les termes de référence de cette expertise[1] et qui stipule que « *L'Asie bénéficie d'un contexte technologique favorable* [en taux d'équipements selon les statistiques de l'ONU]. *On constate toutefois que le numérique est encore peu présent dans les pratiques des établissements membres de l'Agence et la communauté universitaire francophone ; de même la formation à distance y est peu développée : aucune offre de formation à distance en français émanant d'un établissement en Asie n'existe actuellement ; la région ne compte que 14 inscrits en FOAD sur la période 2010-2014 ; seuls trois établissements ont répondu aux appels à projets « Mooc » de l'AUF ces deux dernières années, etc.* ».

Les termes de référence signalent aussi un état de carence « dans les Campus numériques francophones dont les responsables expliquent que les équipements informatiques sont de moins en moins utilisés pour des pratiques individuelles ». La prolifération des technologies numériques mobiles qui devraient normalement constituer un atout important pour le développement de pratiques pédagogiques et de recherche innovante autour des campus numériques francophones, n'a pas été à la hauteur des attentes. Le document évoque un autre détail non moins important qui expliquerait ce paradoxe entre la prolifération des outils technologiques et la diminution des usages lorsqu'il indique qu'en parallèle, et contrairement aux campus francophones, « on observe également dans les formations en langues nationales ou en anglais une dynamique positive d'intégration des TICE et de la distance ». Le document en donne des exemples concrets, notamment celui du programme *ASEAN Cyber University* (ACU) piloté par la Corée du Sud et de ses centres e-learning implantés au Cambodge, Laos, Vietnam et Myanmar ou encore du premier site de Mooc complètement gratuit en langue vietnamienne et des *fablab* mis en place dans la région depuis 2014 sans que la Francophonie y soit impliquée.

Une première hypothèse se dégage de ce constat prémonitoire selon laquelle ce n'est point la technologie qui justifie la démobilisation progressive (voire démotivation) des acteurs locaux pour établir des formes de partenariats francophones pour la formation et la recherche.

Ce n'est pas non plus une question de volonté politique pour améliorer les infrastructures technologiques dans la formation et la recherche par le numérique. Pratiquement tous les entretiens réalisés dans le cadre de cette mission ont démontré la convergence des points de vue et des attitudes ambitieuses exprimées par trois catégories d'acteurs rencontrés :

- des acteurs politiques rencontrés dans les ministères de l'éducation des trois pays qui sont conscients de l'importance du numérique éducatif et des plus-values engendrées par les technologies pour l'éducation. Chacun des trois pays dispose d'un cadre réglementaire et de projets nationaux d'innovation technologique pour l'enseignement et l'éducation par le numérique ;

- des institutions universitaires publiques et privées qui manifestent par la voix de leurs recteurs, présidents et responsables techniques et pédagogiques, leurs convictions profondes pour le numérique éducatif (pour des raisons à la fois de qualité, de compétitivité et d'intérêt économique). Or, vu le modèle de gouvernance plutôt centralisé au niveau des hautes instances de l'État, la majorité des institutions universitaires dans les trois pays en question sont souvent dans l'attente de la promulgation de textes juridiques (décrets, chartes, conventions, directives, etc.) qui leur permettent d'agir et d'adopter des solutions innovantes dans l'enseignement et la recherche ;

---

[1] AUF-BAP (2015). Termes de référence : expertise Formation à distance et outils numériques pour l'enseignement supérieur et la recherche en Asie. Bureau Asie-Pacifique de l'AUF, 10 p. [https://www.auf.org/media/filer_public/f6/c2/f6c281f9-f58a-4656-84a0-75b7c0612348/tdr_expertise_foad_tice_ bap.zip]

- des enseignants-chercheurs relativement peu consultés dans cette expertise, mais suffisamment engagés comme acteurs de terrain pour qu'on puisse prévoir leur points de vue par rapport à l'usage du numérique dans leurs activités pédagogiques et de recherche. Les enseignants et chercheurs, au revenu relativement modeste, auraient inévitablement un rôle décisif dans toute réforme universitaire par le numérique si toutefois des arguments mobilisateurs concrets parviendraient à compenser leur manque à gagner en rejoignant les projets ou programmes de formation ou de développement par le numérique.

## ■ STRUCTURE DE LA PARTIE 2

Dans cette deuxième partie du rapport il sera surtout question de formuler des recommandations et de proposer une feuille de route en vue de programmer des activités dans le cadre du quadriennal de l'AUF 2014-2017.

La partie sera structurée en trois sections principales :

- d'abord un constat général le plus représentatif possible de l'usage des technologies éducatives dans l'enseignement supérieur, de la formation à distance et des CLOMs dans les trois pays visités ;
- ensuite, une série de recommandations en rapport avec les besoins actuels et futurs des établissements universitaires des trois pays ainsi que les points de blocage, les solutions et les opportunités en faveur au développement de la formation à distance et l'intégration des TIC/E dans les pratiques d'enseignement et de recherche ;
- enfin, une feuille de route proposée pour identifier des activités pédagogiques adaptées à la période qui reste de la programmation quadriennale 2014-2017.

> Cette feuille de route est indicative. Elle tentera de fournir des éléments pratiques autour desquels le BAP pourra concevoir une stratégie d'action pour le numérique éducatif francophone durant la deuxième période du quadriennal 2014-2017. Le BAP fera des choix adaptés à sa politique générale, ses moyens budgétaires et ses ressources humaines.

# CONSTAT GÉNÉRAL

Sans rentrer de nouveau dans les détails de chacun des trois pays et de chacune des institutions visitées (ces détails sont accessibles dans la première partie), les visites et les entretiens effectués ont permis de dégager une série de facteurs communs relatifs à l'éducation, l'enseignement supérieur et l'intégration des TIC/E et de la FOAD dans les trois pays visités.

## 1. Faiblesse du cadre législatif des TIC/E et de la FOAD

La faiblesse (voire parfois le vide) législatif en faveur de l'intégration des TIC/E et de la FOAD - comme la non-reconnaissance des diplômes à distance - constitue un handicap commun qu'ont déploré beaucoup de nos interlocuteurs. Il y a en effet une conscience partagée sur l'état précaire de la législation éducative en rapport avec la généralisation des TIC/E. Ce ne sont donc pas les initiatives de déploiement de l'infrastructure numérique qui font défaut, mais plutôt le cadre juridique et légal qui tarde à suivre le rythme des initiatives et des projets afin d'instaurer un contexte formel et officiel propice à des plan d'aménagement éducatifs par le numérique.

Pourtant, les trois pays disposent de beaucoup de textes régissant l'aspect des TIC/E sous forme de circulaires, conventions, accords etc. ; ce qui indique qu'une culture du numérique dans les choix politiques et stratégiques des trois pays dans le domaine de l'éducation ne fait pas non-plus défaut. En revanche, la question qui revient systématiquement chaque fois qu'un nouveau texte est promulgué, est celle de l'adéquation entre les ambitions, parfois démesurées, exprimées dans ces textes, et la réalité du contexte par rapport aux potentialités économiques et technologique du pays et la disponibilité scientifique et culturelle des publics universitaires à croire et adhérer à ces ambitions.

Au Vietnam, si l'on prend juste le cas du Plan stratégique de développement de l'éducation pour la période 2011-2020, l'un des innombrables textes aux ambitions très séduisantes, ce plan prévoit comme objectif en 2020 que « le système d'éducation du Vietnam saura innover fondamentalement et globalement son contenu dans le sens de la normalisation, de la modernisation, de la socialisation, de la démocratisation et de l'intégration internationale. La qualité de l'enseignement sera améliorée globalement, y compris les valeurs éthiques, le style de vie, la pensée créatrice, l'approche pratique, les langues étrangères et les capacités informatiques ». Ce plan prévoit aussi que « la stratégie pour le développement des sciences et de la technologie pour la période 2011-2020 permettra de synchroniser les sciences sociales et les Humanité avec les sciences naturelles et les sciences techniques et technologiques, et de faire de la science et de la technologie une force motrice clé pour répondre aux exigences de base d'un pays industriel moderne ». En 2020, un certain nombre de domaines de la science et de la technologie au Vietnam devraient, selon ce plan, atteindre le niveau de la région de l'ASEAN et celle du monde développé.

Cependant, ces attentes sont jusqu'ici jugées difficiles à atteindre. Dans le *Global Competitiveness Report* 2014-2015, le Vietnam est classé 68e sur 144 pays dans l'indice de compétitivité mondiale (WER, 2014, p.13). Les classements pauvres ont été particulièrement observés dans la qualité de la gestion institutionnelle, dans la disponibilité de projets de recherche d'un impact local et national et dans la disponibilité de programmes de formation à la recherche.

Les préoccupations législatives du système éducatif vietnamien ont été évoquées très récemment pendant un séminaire organisé à Danang le 22 et 23 décembre 2015 à propos du « renforcement d'application de technologie de communication et d'information pour les besoins d'innovation standard et complexe de l'éducation et de la formation », et auquel ont participé des personnalités politiques dont le Vice-ministre de l'Éducation. Il a été question une fois de plus de la fameuse circulaire qui devrait construire et promulguer le référentiel numérique du gouvernement pour l'éducation.

Dans le cas du Cambodge, en regardant de près le texte de quelques circulaires du Ministère de l'Éducation et de la Formation, relatives aux TIC/E (très nombreuses depuis 2003), il est bien clair que les projets sont souvent ambitieux et difficiles à réaliser puisqu'ils sont souvent reconduits à l'identique dans les circulaires suivantes.

Au Laos, les universités attendent aussi un cadre législatif dans lequel le gouvernement définirait une stratégie nationale de réforme universitaire par le numérique qui serait applicable à tous les organismes de l'éducation.

En définitive, bien que l'avancement sur le plan juridique diffère d'un pays à un autre, la conséquence directe à la faiblesse législative dans le domaine des TIC/E et de la FOAD engendre une faiblesse dans l'exploitation optimale d'un potentiel technologique pourtant très présent dans les institutions universitaires.

## 2. Faiblesse de la recherche scientifique

Influencés par le système soviétique, l'enseignement et la recherche ont été historiquement des fonctions distinctes dans les trois pays en question où les recherches sont menées principalement dans les centres de recherche en dehors des établissements d'enseignement supérieur. La séparation institutionnelle de la recherche et de l'enseignement a ainsi engendré un niveau relativement faible de la recherche scientifique dans les universités même si quelques-unes sont aujourd'hui de plus en plus engagées dans des activités de R&D.

Publié en juin 2008, le rapport de la banque mondiale « Vietnam : l'enseignement supérieur et les compétences pour la croissance »[2] - qui peut s'appliquer dans ses grandes lignes aux cas du Cambodge et du Laos - a déjà mentionné ces défaillances et a averti que dans le but de créer un système d'enseignement supérieur de qualité, l'accent doit être mis sur l'élargissement du rôle de la recherche dans les universités. Or, si la recherche ne parvient pas à prendre le pas sur l'enseignement ou du moins suivre son rythme dans le trois pays, c'est en raison de plusieurs facteurs que le rapport de la banque mondiale a bien expliqués :

- *Séparation institutionnelle entre recherche et enseignement* : la majorité du financement gouvernemental pour la recherche continue d'être acheminée vers les instituts de recherche plutôt que les universités et, par conséquent, peu d'universités sont en mesure de fournir des incitations appropriées pour leurs enseignants afin de faire de la recherche-action sur les TIC/E et la FOAD ;

---

[2] World Bank (2010). Vietnam: Higher Education and Skills for Growth. Human Development Department, East Asia and Pacific Region, The World Bank Report No. 44428-VN [https://openknowledge.worldbank.org/handle/10986/7814]

- *Budgets insuffisants pour la recherche* : les universités comptent sur les revenus de la recherche provenant du budget de l'État, avec une quantité négligeable provenant de sources extérieures. Rappelons que le montant du revenu de la recherche d'une université est un indicateur important de sa notoriété scientifique et de sa position dans les classements internationaux des universités ;

- *Carence de personnel enseignant qualifié* : un grand pourcentage du personnel académique dans les universités des trois pays accède au rang de professeur sans pour autant détenir un titre de doctorat. Ce manque d'universitaires de haut niveau dans l'enseignement supérieur est souvent le résultat des coûts exorbitants pour les budgets de l'État ou des institutions pour avoir des formations doctorales ;

- *Faiblesses de la production scientifique* : parce que les activités de recherche de bonne qualité engagent un facteur temps important, alors qu'il n'y a pas de valorisation à la hauteur des engagements réalisés, la plupart des professeurs dans les universités publiques préfèrent enseigner ou accomplir des tâches administratives, voire réaliser des activités rémunérées dans le domaine privé ;

- *Faiblesse des salaires* : bon nombre des problèmes liés à la qualité du personnel enseignant ont trait à la faiblesse de leurs salaires et des procédures lourdes pour les promotions qui ne récompensent pas suffisamment le travail académique. Le personnel universitaire dans les établissements publics est considéré comme fonctionnaire dont la nomination, la promotion, la rétrogradation ou la résiliation dépendent de l'institution, sous contrôle de la fonction publique ;

- *Faiblesse des infrastructures technologiques* : les accès aux ordinateurs et la connexion à Internet par les étudiants restent aussi très faibles dans la majorité des établissements de l'enseignement supérieur public ;

- *Précarité des conditions de travail dans les universités publiques* : les conditions dans les établissements d'enseignement supérieur découragent assez souvent la participation des professeurs dans la recherche universitaire. C'est souvent dû à des charges élevées d'enseignement, à un grand nombre d'étudiants à suivre, au manque de conditions appropriées de travail (beaucoup ne disposent pas de bureaux ou de lieux pour mener des recherches) et à l'absence d'incitations financières pour s'engager dans une recherche de longue haleine…

## 3. Démotivation des enseignants à la formation et à la numérisation des cours

Ce critère est fondamentalement crucial et complexe pour plusieurs considérations : économiques, législatives et déontologiques :

- *Raisons économiques* : le facteur économique est hautement important dans la définition d'une politique de réforme de l'enseignement et de la recherche par le numérique éducatif. Si tous les entretiens effectués ont soulevé la question économique comme facteur déterminant des contraintes que rencontrent les projets de réforme, tous fournissent aussi les mêmes raisons, notamment les salaires trop bas et trop peu incitatifs. Les enseignants seraient moins bien payés que la moyenne des salaires de l'ensemble des secteurs économiques. Aussi, exercent-ils souvent un autre emploi à mi-temps et, par conséquent, sont régulièrement absents dans les classes. En somme, la pauvreté constitue un facteur déstabilisant pour tout le système éducatif, affectant le recrutement des enseignants, la qualité des infrastructures scolaires, la qualité du curriculum et le niveau de ressources affectées au système éducatif ;

- *Raisons juridiques* : le second critère justifiant la démobilisation des enseignants est législatif et réglementaire. Comme l'ont confirmé plusieurs interlocuteurs, aucune loi ne règle les pratiques des enseignants et ne définit des procédures pour la distribution systématique de leurs cours malgré la conviction collective de l'importance de cette démarche au profil de la réforme de l'enseignement. Même dans les pratiques de beaucoup d'universités qui ont plus d'autonomie dans leurs gestions internes, celles-ci ne parviennent pas à en faire un processus organisé et systématique ;

- *Raisons déontologiques* : le problème déontologique est évoqué sous deux aspects. Le premier est la rétention des données de peur de se faire voler ses idées ou son travail, ce qui laisse deviner une faiblesse de culture liée aux principes des « Droits d'auteur » et de la « propriété intellectuelle ». D'où la nécessité de penser à renforcer ces notions essentielles chez les enseignants universitaires. Le deuxième aspect est relatif à la qualité scientifique des cours dispensés. Des enseignants évitent souvent d'exposer publiquement les contenus de leurs cours de peur de dévoiler des opérations de plagia, ce qui laisse deviner une autre faiblesse de culture liée aux deux pratiques scientifiques de la « citation » et de « l'évaluation par les pairs ». La faiblesse de l'activité de recherche scientifique, comme c'est décrit précédemment, et sa dissociation de l'action pédagogique explique ce type de comportement réfractaire. En plus des programmes de formation sur les techniques, une formation sur la culture du numérique, notamment en lien avec les logiciels libres, le libre accès, les droits d'auteurs et la propriété intellectuelle est vivement recommandée dans tout programme de formation sur les TIC/E et la FAOD.

## 4. Différences en équipements TIC/E et en ressources numériques

L'écart souvent très important en équipement et en ressources numériques est surtout observé entre les universités publiques et privées. Bien que soumises dans la majorité des cas aux directives ou à la tutelle des Ministère de l'éducation, les universités privées bénéficient d'une autonomie budgétaire issue des frais d'inscription et des financements de projets de coopération (internationale ou avec des entreprises nationales) qui leur permettent de disposer de plus de moyens financiers.

Les universités publiques – comparées entre-elles - ne sont pas non plus au même niveau d'avancement du point de vue des TIC/E et de la FOAD. Les écarts sont souvent déterminés en fonction des partenariats établis avec des acteurs régionaux et internationaux. C'est le cas, par exemple des universités et instituts qui disposent ou non de centre e-Learning ou de laboratoire multimédia fournis dans le cadre de programmes régionaux comme l'ASEAN Cyber University (ACU), l'ASEAN University Network (AUN) ou l'Agence Coréenne de Coopération Internationale (KOICA). À des degrés moindres, certaines universités reçoivent des équipements et échangent des ressources numériques avec des universités et des organismes internationaux en Europe, en Australie, au Japon ou aux États-Unis d'Amérique.

## 5. Carence des contenus en ligne

L'insuffisance des contenus pédagogiques numériques en ligne est également une constante dans tous les bilans relatifs à la FAOD dans les trois pays visités. De façon générale, les raisons sont les mêmes, à savoir le manque de motivation des enseignants (raisons expliquées plus haut), le manque de personnel qualifié pour entreprendre la numérisation des contenus, la faiblesse du cadre législatif instituant la restitution ou le dépôt systématique des contenus des cours, etc.

Parmi les conséquences de l'absence d'une stratégie de mise en ligne des cours, les réservoirs ou les portails de ressources pédagogiques (libres ou non) sont très rares, voire quasiment inexistants. Certaines universités, comme l'Université Ho Chi Minh de Technologie, supprime les cours en ligne déposés par les enseignants sur Moodle au bout de six ans alors qu'il aurait été plus utile, sous réserve de régler la question des droits d'auteurs, de procéder à l'archivage permanent de ces ressources et d'en constituer un patrimoine éducatif institutionnel ou national.

Plusieurs autres cas de figure ont été identifiés et tous dénotent de l'absence d'une culture de capitalisation, de mutualisation et de conservation d'un patrimoine numérique éducatif important. Il va sans dire aussi que l'absence d'une culture normative aussi bien dans l'indexation des ressources numériques que dans leur référencement sur les réseaux pour en faciliter la restitution, accentue cette défaillance. Dans ce champ d'expertise, l'expérience francophone de l'AUF, notamment le projet IDNEUF[3] (Initiative pour le Développement du Numérique Éducatif Universitaire Francophone) et le référentiel des compétences TIC/E[4] de l'IFIC seraient d'une grande utilité pour mettre en place des programmes de formation destinés à la communauté universitaire dans les trois pays.

## 6. Moodle comme CMS fétiche en Asie pacifique

Le constat commun aux trois pays CLV est que la FOAD dans sa majeure partie est assimilée à une distribution de cours en ligne. C'est la variation dans la nature technique des cours et de leur structuration qui crée la différence entre un système FOAD et un autre. Certains dispositifs comme ceux de l'ACU à l'Université de Hanoï des Sciences et des Technologies (HUST) ou à l'Institut de Technologie du Cambodge à Phnom-Penh (ITC) procèdent par voie d'enregistrements vidéo pour distribuer des cours, d'autres, comme l'Universités des technologies de Ho Chi Minh, intègrent les cours en formats PDF aux services numériques de leurs portails institutionnels.

Le choix de Moodle comme plate-forme pédagogique déployée un peu partout dans les pays CLV est un signe à double sens. D'abord c'est un marqueur d'une culture FOAD généralisée qui opte pour un CMS libre et ouvert pour dispenser des contenus pédagogiques en ligne. Cela dénote d'une tradition d'enseignement à distance bien ancrée dans la conscience collective éducative du pays. En revanche, à part quelques exceptions, Moodle est utilisé un peu partout comme une plate-forme de diffusion de cours souvent sans activités pédagogiques associées comme l'accompagnement des apprenants dans des parcours d'apprentissage individualisés ou des activités de tutorat. D'un LCMS (son identité réelle), Moodle est exploité comme CMS, le « L » de « Learning » étant délaissé pour compte.

## 7. Francophonie et diversité linguistique

L'aspect linguistique en rapport avec la langue française a été un élément constant dans tous les entretiens. La majorité convient que les départements de français dans les universités sont désormais presque les seuls à utiliser la langue française, toutes les autres disciplines étant dispensées en anglais ou en langues nationales. Certains considèrent que la politique de bourses suivie par les ambassades de France devrait être consolidée par un réseau de formation de la langue française en vue de créer une masse critique de francophiles capables de constituer des réseaux d'influence et de collaboration. Certains critiquent, à tort ou à raison, les actions francophones sur deux points. Le premier est la focalisation de la Francophonie dans ses projets de coopération sur les sujets culturels et linguistiques alors que la Francophonie peut évoluer à travers d'autres créneaux disciplinaires dans d'autres domaines des sciences. Le deuxième point est celui des formations réalisées par des acteurs francophones (Ambassades, AUF, OIF) qui, pour certains, ne devraient pas se limiter à l'usage du français comme langue de communication, mais devraient parfois se faire dans les langues partenaires locales. La mobilisation pour la Francophonie, selon eux, n'est pas exclusivement dans l'usage de la langue mais aussi dans la prestation des services. L'intérêt pour une langue commence d'abord par l'intérêt pour ce que cette langue véhicule et non par la forme comment elle l'exprime. En créant l'intérêt pour un contenu ou un service communiqué dans la langue du locuteur, celui-ci chercherait ensuite à s'approprier la langue source pour avoir accès en toute autonomie à ce contenu ou service qui l'intéresse.

---

[3] Cf. http://ific.auf.org/sites/default/files/POUR%20UN%20PORTA@@Synth%C3%A8se%20portail%20Francophone.pdf
[4] Cf. http://ific-auf.org/transfer/le-r%C3%A9f%C3%A9rentiel-tictice

Une autre remarque non moins importante nous est parvenue de l'un de nos interlocuteurs. Beaucoup de recteurs d'universités et de hauts responsables académiques et politiques sont d'une génération qui a connu l'époque glorieuse de la langue française dans la région (i.e. au sein de la Fédération indochinoise). Ils ont été formés à la culture et à la langue française et y portent un attachement particulier. Ces décideurs sont en train de prendre leurs retraites pour laisser place à des jeunes responsables moins francophones et plus orientés vers la mondialisation. Aussi, la Francophonie gagnerait-elle beaucoup à préparer une génération de relève. Elle devrait trouver les mécanismes nécessaires pour se régénérer non pas par des processus d'apprentissage du français comme langue étrangère uniquement, mais comme processus qui tient compte du français comme langue partenaire de travail, d'enseignement et de recherche scientifique.

Nous proposons ici une démarche que la Francophone pourrit envisager dans sa stratégie linguistique en Asie-pacifique : la formation des formateurs dans les langues locales. Il faut bien reconnaître le poids très lourd de la concurrence linguistique face à l'hégémonie mondiale de l'anglais ; une réalité que l'OIF admet clairement dans une intervention intitulée « Politique intégrée de promotion de la langue française » présentée au Sommet de Kinshasa en octobre 2012 : « L'anglais s'impose de plus en plus comme la langue quasi unique de la modernité, de la technologie, de la norme et de l'économie, ce qui relègue toutes les autres langues, y compris le français, au second plan [...] De plus, avec la montée des pays émergents, une nouvelle structuration mondiale des échanges linguistiques est en train de se forger. Cette mutation invite à agir en priorité sur les facteurs qui conditionnent le statut international du français, mais aussi à imaginer sa place et son rôle dans le nouveau rapport qui s'instaure peu à peu entre les langues d'usage international » (OIF, 2012)[5].

À ce titre, la Francophonie gagnerait à étudier la possibilité de réviser deux types de frontières linguistiques : l'usage (total ou partiel) du français comme règle d'adhésion, et l'usage du français comme langue quasi exclusive de formation.

Sur le premier point, la Francophonie pratique une exclusion sur la base de la langue d'enseignement, totale ou partielle, alors que nous avons été témoin du cas exceptionnel de l'HUST à Hanoï qui, en raison des contraintes de l'employabilité de ses ressortissants, pratique un enseignement exclusivement en anglais avec des enseignants français et des nationaux diplômés d'universités françaises. Au lieu de fermer la porte à des institutions potentiellement utiles, il faudrait peut-être revoir la politique francophone dans ce sens pour trouver des mécanismes permettant de profiter de l'appui d'institutions qui ne sont pas en Francophonie mais qui disposent d'un potentiel et d'une marge importante de pénétration dans le tissu social d'un pays. Ceci pourrait-il ainsi ouvrir la voie vers plus d'alternatives d'enseignement du français au sein des populations locales.

---

[5] Cf. http://www.francophonie.org/IMG/pdf/politique_integree_de_promotion_de_la_langue_francaise.pdf

Sur le deuxième point, il faudrait éventuellement aussi modérer le critère du français comme langue quasi exclusive de formation et prendre l'exemple d'expériences internationales qui, en s'inscrivant dans la formation à distance multilingue enregistrent des taux immenses de participation. Bien que de nature différente que les formations proposées par l'AUF, le programme Global Learning des Nations-Unies pour la lutte contre la drogue et le crime (UNODC) est conçu pour offrir des formations en ligne et hors ligne dans le monde entier. Sa plate-forme en ligne[6] dénombre plus de 3200 utilisateurs. Les cours de formation sont sur mesure, élaborés par l'UNODC en collaboration avec des experts internationaux. Ils correspondent directement aux besoins des États membres.

Proposer des formations en ligne dans les langues nationales sur des sujets touchant directement les populations locales a été également l'une des recommandations de la 11e édition du séminaire régional de recherche-action sur la question des usages du français dans la région de l'Asie-pacifique. Ce séminaire, organisé à Hô-Chi-Minh-Ville du 23 au 25 novembre 2015 par le Centre régional de l'OIF pour l'Asie et le Pacifique (Crefap)[7], a permis à 82 enseignants chercheurs en provenance de sept pays francophones (Belgique, France, Cambodge, Canada, Laos, Thaïlande, Vietnam, Vanuatu) plus la Malaisie, la Chine et Singapour, de discuter de la position du français dans la région à travers ses représentations, ses pratiques et les politiques linguistiques des États. Les constats ont fait part d'une situation extrêmement disparate avec des pays où le français se porte bien – la Chine, la Thaïlande – et une inquiétude face au recul du français au Laos, au Cambodge et au Vietnam. Au vu des mécanismes d'enseignement et des cursus proposés, il est apparu que le français ne souffrait pas d'un manque de cohérence dans son enseignement, mais qu'il n'était pas suffisamment associé, dans les représentations des jeunes et de leurs parents, aux filières susceptibles d'offrir des débouchés professionnels.

Afin de réorienter les actions, les participants ont proposé trois nouvelles directions à suivre :

- La contextualisation territoriale : des formations adaptées aux besoins des régions plutôt qu'aux pays ;
- La professionnalisation : l'articulation nécessaire à opérer entre le français et le champ professionnel par des curricula et des actions de terrain adaptés ;
- Le plurilinguisme : les formations trilingues (langue nationale, français, anglais) deviennent une exigence globale et permettent la survie des langues.

Pour conclure notre conception sur le plan du positionnement du français dans le tissu social des pays asiatiques, nous considérons qu'il est primordial de commencer par prévoir un renforcement de la langue française dans un premier cercle d'ancrage, celui d'un fort partenariat avec les langues locales ; Ensuite, il faudrait passer à un cercle de motivation par le français dans le domaine de l'enseignement et de la recherche scientifique, pour aboutir enfin à un cercle de rentabilité où la langue française augmenterait son quotient d'employabilité comme langue de travail.

Les coréens pratiquent judicieusement cette stratégie dans la planification des programmes de l'ACU. Après avoir initié un cycle formation en 2012 sur des techniques diverses en langue anglaise, ils ont programmé une formation gratuite en 2014 sur la langue coréenne. Ainsi dilué dans une programmation à plusieurs facettes, l'accès à une langue autre que la langue nationale passerait beaucoup plus facilement auprès des partenaires locaux.

---

[6] Cf. https://www.unodc.org/elearning/frontpage.jsp
[7] Cf. http://www.francophonie.org/Quels-usages-du-francais-dans-la.html

Les coréens ont pu faire de manière plus subtile : dans le programme de l'ACU, plusieurs partenaires des pays CLMV ont été invités en Corée pour une formation technique en utilisant l'anglais comme langue pivot. Des formateurs coréens se sont également déplacés pour former sur place dans les universités partenaires. Or, même si l'anglais joue le rôle de *lingua franca* dans ces échanges et ces formations, nous avons eu des témoignages d'enseignants qui ont cherché volontairement à apprendre le Coréen pour mieux profiter des ressources produites dans le cadre de l'AUN et de KOICA.

Pour la Francophonie, ce scénario pourrait être envisagé autant dans le cadre de la formation des formateurs que dans l'appui aux universités pour le développement de leurs offres de formation auxquelles le BAP participe activement dans le cadre du quadriennal 2014-2017.

## 8. POUR RÉSUMER !

Admettons toutefois qu'en l'absence de directives générales et harmonieuses pour l'introduction de la FOAD et des TICE dans les programmes de réforme de l'enseignement supérieur des pays visités, ce domaine est tributaire de contextes et de conditions différentes et variées. La revue de la littérature et les visites sur site ont démontré des divergences dans la conception et les modes d'usage des TICE et de la FOAD. Mais toutes convergent vers un constat relativement commun :

- La carence de contenus pédagogiques par rapport à la taille des institutions et du nombre de lectorat ;
- La carence de professionnels qualifiés en FOAD. La majorité des enseignants n'ont pas la formation adéquate pour utiliser les outils de la FOAD ;
- La fragilité de l'approche pédagogique fondée sur la dynamique de groupes et la validation des compétences par logique de projets ;
- La carence des budgets adéquats pour mettre en place des systèmes FOAD opérationnels ;
- La rareté des opportunités de formation d'experts en FOAD ;
- La précarité des infrastructures FOAD ;
- La précarité de la recherche scientifique censée appuyer la culture numérique autour de la FOAD.

En général, les rapports et les analyses sur le sujet décrivent des exemples parfois très différents et rendent des témoignages qui vont de l'exaltation pour des expériences privées dans la formation professionnelle (*vocational training*) au relativisme de la FOAD comme levier significatif de la réforme de l'enseignement supérieur dans la région. Le paysage de la FOAD est, de ce fait, très disparate et dénivelé entre les secteurs privés et publics, entre la formation professionnelle et l'enseignement scolaire. Les gouvernements ont tendance à donner la priorité aux enseignements primaires et secondaires, en contrepartie d'une plus large autonomisation du supérieur, ce que des experts qualifient de grave problème pour la FOAD en raison du manque de normes nationales uniformisées dans cette méthode de formation. Chaque établissement a sa propre méthode pédagogique et ses propres manuels. La gestion des effectifs d'enseignants est aux abonnés absents, avec parfois pour conséquence un manque de formateurs. De plus, pour espérer terminer une formation à distance, les apprenants doivent disposer de quatre grandes qualités : motivation, autonomie, discipline et autogestion, des qualités que les jeunes ne sont pas habitués à avoir et à développer.

# RECOMMANDATIONS

À l'issue du ce rapide tour d'horizon exploratoire du numérique éducatif dans les trois pays CLV, nous sommes amenés à fournir une série de recommandations que nous estimons utiles pour un numérique éducatif francophone dans la région de l'Asie pacifique ; un numérique éducatif qui serait entrepreneur, innovant et compétitif. Le numérique éducatif véhicule des enjeux stratégiques d'ordre politique, économique, culturel, linguistique, scientifique et technique sur un fonds de valeurs humaines et sociales profondes. Cependant, ce rapport étant prévu pour faire l'état des lieux sur les aspects relatifs au déploiement des TIC/E et de l'usage de la FOAD au Cambodge, Laos et Vietnam, il sera limité à une série de recommandations qui devrait aider à la définition d'une feuille de route pour la programmation des activités du Bureau de l'AUF dans la région (BAP) pour la période 2016-2017.

Cette feuille de route devrait constituer un socle d'appui à la réalisation des objectifs de la mission « Numérique éducatif et services à la communauté universitaire francophone du bureau régional Asie-Pacifique » qui ambitionne d'« accompagner l'évolution et la diversification des pratiques pédagogiques et de recherche des établissements membres de l'AUF dans la région, à structurer un réseau d'infrastructures et d'expertise et à renforcer les outils et services numériques mis à la disposition de la communauté universitaire francophone.».

Or, les objectifs de cette mission ne sauront être réalisés sans une transversalité et une complémentarité avec les autres missions du BAP articulées dans une architecture de programmation quadriennale structurée en mission – axe – projet – activité.

Pour tenir compte de cette articulation des missions et de leurs objectifs, et sans préjuger à ce jour des choix qui seraient pris en compte par les acteurs francophones en charge de la rénovation universitaire par le numérique, nos recommandations dans le cadre de la dite mission pourrait être structurées sous les trois axes suivants :

1) Un premier axe relatif au développement des ressources numériques éducatives, leur gestion, organisation et distribution ;
2) Un deuxième axe relatif à la formation et le renforcement des capacités autour du numérique éducatif ;
3) Un troisième axe relatif à l'appui à la recherche scientifique autour du numérique éducatif.

Chacun de ces axes peut ensuite être décliné en autant de sous axes, projets et activités que le permettraient les moyens humains et logistiques disponibles. Une répartition des rôles/tâches peut être envisagée au sein d'un dispositif régional d'appui au numérique éducatif francophone au sein d'un réseau de campus numériques francophones et campus numériques francophones partenaires dans la région Asie pacifique.

Cette structuration est proposée en tenant compte de deux sources d'inspiration essentielles : d'abord l'état des lieux que nous avons pu établir lors de nos visites aux trois pays concernés, mais aussi la politique du numérique éducatif conduite actuellement au sein de l'AUF, de l'IFIC et du BAP, structures francophones lourdement engagées dans la réalisation de projets de rénovation par le numérique éducatif à l'échelle de la Francophonie, dans le monde et dans les régions.

# 1. DÉVELOPPEMENT DE RESSOURCES NUMÉRIQUES ÉDUCATIVES

La production des ressources éducatives libres est au cœur de la vision 2014-2017 de l'AUF. La production des RELs s'inscrit dans l'appui francophone aux logiciels libres, à l'accès ouvert et aux licences libres du type *creatives commons*. La proposition de cet axe dans la liste des recommandations sera argumentée par des éléments issus à la fois de l'état des lieux observé dans les pays visités (CLV) et aussi dans la politique du numérique éducatif en cours de l'AUF.

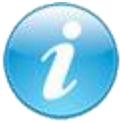
La promotion des Ressources Éducatives Libres (REL) en Francophonie s'inscrit dans l'optique de la déclaration de Dakar sur les Ressources Éducatives Libres promulguée en mars 2009 puis celle de Paris de l'UNESCO adoptée en 2012 lors du congrès mondial des Ressources éducatives libres auquel l'AUF et l'OIF avaient pris part. Ces deux déclarations engagent le BAP à promouvoir les RELs et les logiciels libres.

## 1.1. État des lieux dans les pays CLV

Dans notre étude de l'état des lieux nous avons soulevé comme problème principal du numérique éducatif dans les trois pays visités, l'insuffisance de contenus éducatifs numériques développés et partagés sur les réseaux des universités.

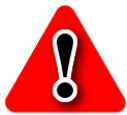
L'insuffisance des contenus pédagogiques numériques en ligne est une constante dans tous les bilans relatifs à la FOAD dans les trois pays visités. Tout dénote de l'absence d'une culture de capitalisation, de mutualisation et de conservation d'un patrimoine numérique éducatif important. Il va sans dire aussi que l'absence totale d'une culture normative aussi bien dans l'indexation des ressources numériques que dans leur référencement sur les réseaux pour en faciliter la restitution, accentue cette défaillance.

Nous avons aussi évoqué les raisons inhérentes à cette carence qui n'est point un problème d'inexistence de ressources éducatives libres (REL) - d'autant plus que les enseignants produisent réellement des contenus pour leurs cours - mais plutôt une réticence à les partager par manque de compensation financière et/ou de valorisation scientifique. Cette attitude de rétention, stimulée également par une carence réglementaire pour imposer la remise et le partage des REL sur les dispositifs et les portails des insultions universitaires (Moodle comme CMS ubiquitaire), connait plusieurs conséquences négatives, notamment :

- inaccessibilité à une grande partie du support didactique, rédigé souvent en langues locales ;
- impossibilité de création d'archives patrimoniales éducatives ;
- un gaspillage important dans la duplication des contenus utilisés dans des formations similaires dans des instituions du même domaine. Il nous a été donné de constater dans les trois pays l'absence de consortium universitaires, de regroupements thématiques ou disciplinaires entre institutions du même domaine, comme les deux universités de médecine à Ho Chi Minh Ville, l'une sous tutelle de la ville et l'autre sous tutelle du Ministère de la santé ;
- des contenus de cours pourtant obsolètes, continuant à être dispensés sans contrôle de qualité ;

Par contre, ce n'est pas uniquement le problème de rétention qui cause problème par rapport aux ressources éducatives numériques. Même quand ces ressources sont disponibles et partagées sur des sites institutionnels, elles manquent souvent de conformité - dans leur conception et leurs modes de référencement - avec les normes et les standards internationaux qui pourraient faciliter leur recherche, récupération et mutualisation. Un travail de fond est à prévoir sous cet angle pour améliorer l'état des ressources éducatives numériques produites et à produire. Cette ligne devrait entrer dans un schéma de démarche qualité de l'enseignement supérieur.

## 1.2. Activités francophones de développement de RELs en Asie pacifique

Plusieurs actions de formation sur la « Conception, développement et utilisation d'un cours en ligne » ont été déjà programmées dans la région du BAP dans le cadre des formations Transfer de l'Agence universitaire de la Francophonie. Ces formations s'adressant à l'ensemble des enseignants et responsables scolaires et universitaires appelés à participer à la mise en place de dispositifs d'EAD au sein de leurs institutions, permettaient l'acquisition de compétences conceptuelles et techniques pour réaliser un cours multimédia-interactif, pour une mise en ligne ou pour un enseignement.

Or, depuis lors, la notion de ressources pédagogiques a évolué pour prendre es dimensions nouvelles tant sur le plan de la conception (indexation, référencement) que celui de la diffusion et communication (ressources libres). On parle désormais plus de ressources éducatives libres (RELs) que de cours en ligne.

Plusieurs actions clés ont contribué à ce changement :

- d'abord la « Déclaration de Dakar sur les Ressources Éducatives libres » (REL)[8] promulguée en 2009 à l'initiative du Bureau Régional pour l'Éducation en Afrique de UNESCO, de l'Organisation Internationale de la Francophonie (OIF), et de l'Agence universitaire de la Francophonie (AUF) ;

- puis la déclaration de Paris de l'UNESCO adoptée en 2012 lors du congrès mondial des Ressources éducatives libres auquel l'AUF et l'OIF avaient pris part ;

- ensuite la première session de formation dédiée aux professionnels de l'enseignement et de l'éducation, organisée à Abidjan, Côte d'Ivoire du 05 au 09 octobre 2015 qui a été le déclencheur d'un mouvement francophone d'appropriation des RELs et de leur mode d'accès libre.

Aussitôt, une formation sur les RELs a pris forme dans la région Asie pacifique conduite par le Bureau Régional Asie Pacifique de l'Organisation Internationale de la Francophonie, en collaboration avec la Direction Éducation et Jeunesse et la Direction de la Francophonie Numérique. La toute première formation de ce type a été celle de Hanoï du 16 au 18 décembre 2015intitulé « Production de ressources éducatives libres francophones ». Cette formation était destinée à des professionnels de l'éducation (didacticiens, pédagogues, professeurs) d'Asie et du Pacifique des pays membres de l'OIF (Cambodge,

---

[8] AUF, OIF, UNESCO. Déclaration de Dakar sur les Ressources Éducatives libres (REL), 2015. [https://www.auf.org/media/IMG2//pdf/REL-Declaration_de_Dakar-5_mars_2009.pdf]

Laos, Vanuatu et Vietnam) et leur a notamment permis d'aborder l'approche des logiciels libres et ouverts et l'utilisation des outils libres de création de ressources éducatives libres9.

Ce mouvement est à ses débuts dans la région comme d'ailleurs partout dans les autres régions francophones. L'AUF y tient beaucoup et le recommande en appui au projet du portail francophone des ressources éducatives libres en construction ans le cadre du projet IDNEUF.

### 1.3. Politique numérique francophone : le projet IDNEUF

La question du rassemblement, du référencement, de l'indexation et de l'accès transparent aux ressources pédagogiques en ligne est au cœur de la politique du numérique éducatif francophone. Pour cette raison, la décision de création d'un futur portail francophone commun de REL a été prise par les ministres de l'enseignement supérieur des pays francophones réunis à Paris le 5juin 2015. Cette réunion a permis de dresser l'état des lieux des ressources numériques dans l'espace universitaire francophone, d'évoquer les besoins et le rôle des universités et de réfléchir à une meilleure mutualisation des ressources existantes. Les ministres ont adopté à cet égard une déclaration commune[10] annonçant des mesures à suivre dans le cadre d'une Initiative de Développement Numérique de l'Espace Universitaire Francophone (IDNEUF).

Le portail de RELs est l'une des recommandations confiée à l'AUF. Une étude exploratoire intitulée « Étude sur les portails et agrégateurs des ressources pédagogiques universitaires francophones en accès libre » a été réalisée par l'auteur de ce rapport pour étudier la typologie du portail à mettre en place[11].

Le 17 juin 2016, une équipe d'experts a déjà mis en phase de production ce méta-portail qui recense près de 40 000 ressources numériques présentée lors de la deuxième rencontre des ministres francophones de l'Enseignement supérieur, à Bamako (Mali) en juin 2016. Baptisé IDNEUF, ce "méta-portail" propose un point d'accès unique vers des ressources pédagogiques mutualisées provenant d'une douzaine d'universités francophones.

IDNEUF permet à tous les étudiants, aux enseignants, aux chercheurs et aux autres publics en formation (ingénieurs, techniciens, personnel administratif, etc), d'accéder gratuitement à des milliers de ressources de tous types : cours en ligne, supports de cours, exercices, articles scientifiques, scénarios pédagogiques, etc. grâce à un moteur de recherche pointu qui permet aux utilisateurs de trier les ressources par pays, discipline ou type de ressources.

Concrètement, dans le projet IDNEUF, chaque université francophone peut (et devrait) contribuer pour archiver et diffuser son patrimoine numérique qu'il soit documentaire, scientifique ou pédagogique, conformément à sa politique d'établissement, et avoir accès aux documents des établissements qui participent au portail francophone. Une pareille construction de portail de REL permettrait d'améliorer le partage de ressources et de faciliter la création de nouvelles connaissances, tant sur le plan pédagogique que scientifique et culturel :

- Les enseignants peuvent s'appuyer sur les cours, les travaux pratiques, les exercices, les applications, etc. déjà réalisés par leurs collègues pour créer de nouveaux contenus plus riches, plus clairs et plus innovants ;

---

[9] OIF. "Asie-Pacifique : production de ressources éducatives libres francophones". [http://www.francophonie.org/Asie-Pacifique-REL-46638.html]

[10] Cf. http://www.francophonie.org/IMG/pdf/declaration_ministres_enseignement_ressources_numeriques.pdf

[11] Cf. http://ific.auf.org/sites/default/files/%40%40Etude%20des%20portails-francophone-V02.pdf

- Les étudiants peuvent découvrir de nouveaux éclairages sur les concepts qu'ils doivent apprendre pour préparer leurs examens à partir d'un plus grand nombre d'exercices, ces derniers ayant été finement et richement indexés par des professionnels compétents ;
- Les chercheurs peuvent accéder facilement à l'ensemble de la production scientifique des établissements partenaires, cette production ne se limitant pas aux articles publiés dans les grandes revues, puisqu'elle comprend aussi les actes de colloques, les chapitres dans les publications collectives, les thèses, les mémoires, etc. Ils peuvent ainsi mieux cibler leurs propres recherches ou trouver de nouvelles idées ;
- Les établissements peuvent mieux suivre les évolutions dans les domaines qui les intéressent.

### 1.4. Recommandations pour le BAP

Le méta-portail donne aujourd'hui accès à près de 40000 ressources numériques d'enseignement produites surtout par la France, le Québec, la Tunisie, le Maroc, le Cameroun, le Sénégal, le Liban, le Burkina Faso, l'Égypte, le Mexique.

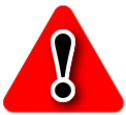

Aucun pays de la région Asie pacifique n'est encore parmi la liste des dépositaires de RELs dans le portail IDNEUF.

Le BAP, comme tous les bureaux régionaux de l'AUF, est donc appelé à alimenter la base des ressources du méta-portail IDNEUF. Il s'agit d'un élément déterminant pour l'AUF en tant que maître d'ouvrage pour démontrer que le méta-portail est bien opérationnel et utile à l'échelle de la Francophonie. C'est la raison pour laquelle l'AUF appelle les responsables des CNFs et tous ceux qui peuvent y contribuer, à se mobiliser y compris en allant dans les universités partenaires et en sensibilisant des enseignants-chercheurs dans leurs régions pour qu'ils mettent à disposition des ressources pédagogiques numériques pour ce méta-portail. Ces ressources peuvent dans cette phase expérimentale être indexées ou non, car l'indexation normalisée est prévue comme phase d'un développement avancé nécessitant des référentiels normatifs de métadonnées pédagogiques appropriés sur lesquels l'IFIC travaille actuellement.

À l'issue de cette présentation sommaire du projet IDNEUF et du portail des Ressources Éducatives Libres, des recommandations pratiques peuvent toujours être proposées au BAP pour confirmer sa participation au projet :

- Faire une étude de l'existant pour inventorier les acteurs institutionnels producteurs potentiels de ressources éducatives numériques dans la région Asie pacifique ;
- Mettre en œuvre une politique de conventions avec les institutions universitaires francophones de la région pour harmoniser les procédés de collecte des REL ;
- Mener une étude exploratoire au sein des universités partenaires de la région pour identifier des personnes ressources productrices de ressources pédagogiques et leur proposer des conventions de cession de droit d'auteur de leurs ressources pédagogiques numériques ;

- Établir un inventaire typologique et quantitatif des ressources pédagogiques rassemblées (genre, volume, niveau et granularité de chaque type de ressource) qui doit remonter depuis chaque institution partenaire vers l'équipe de pilotage locale puis vers le portail des REL francophone ;
- Identifier tout autre type de supports pédagogiques francophones éparpillés et mal exploités dans les institutions universitaires francophones ;
- Envoyer au portail des REL francophones des ressources pédagogiques numériques propres au BAP (projets de recherche, contenus des ateliers de formations, politique de numérisation, etc.) ;
- Préparer des mesures d'accès aux des ressources pédagogiques du portail francophone à partir des sites institutionnels du BAP ;
- Envisager une équipe permanente de suivi du portail qui impliquerait a minima un documentaliste ou un spécialiste e-learning pour indexer les REL et les transmettre au portail IDNEUF ;
- Former une (des) personne(s) ressource(s) dans les bibliothèques et centres de documentation des institutions partenaires pour assurer la collecte et l'indexation des ressources éducatives disponibles ;
- Élaborer un manuel de procédures qualité au profit des partenaires universitaires pour rappeler les directives fondamentales concernant la production harmoniée des ressources et leur traitement comme des ressources interopérables et partageables en ligne ;
- Développer sur les longs termes un vocabulaire multilingue (français/langues locales) pour le besoin des opérations d'indexation/recherche des ressources pédagogiques ;

Il est clair que tout travail de collecte des REL devrait se faire sur la base de conventions, accords et règlements de partenariats solides qui définissent clairement les obligations et les droits de chacun des acteurs concernés. Il importe de définir clairement les politiques que les parties prenantes s'engageront à respecter, d'obtenir un commun accord et une compréhension commune sur la propriété intellectuelle et sur la responsabilité de chacun, ainsi que sur l'édition et la contribution des ressources. Les politiques sont importantes pour clarifier les responsabilités des acteurs, faciliter la planification des opérations et choisir/construire les outils d'assistance. Cette politique doit expliquer comment établir une procédure suivant laquelle chaque ressource est évaluée avant d'être déposée dans le portail. La politique du portail précise en général les exigences pour qu'une ressource soit acceptée. Rappelons enfin qu'un rapport plus détaillé explique mieux les points évoqués sous cette rubrique relative au portail IDNEUF des ressources pédagogiques francophones libres (Ben Henda, 2015)[12].

## 2. RENFORCEMENT DES CAPACITÉS AUTOUR DU NUMÉRIQUE ÉDUCATIF

Non moins importants que le développement et le partage des ressources, l'enseignement et la formation sur les technologies éducatives TIC/E et la FOAD participent à la fois de la bonne gouvernance d'une stratégie de réforme universitaire par le numérique éducatif et du renforcement des capacités numériques pour l'insertion professionnelle.

---

[12] BEN HENDA Mokhtar, « Étude sur les portails et agrégateurs des ressources pédagogiques universitaires francophones en accès libre ». IFIC, AUF, septembre 2015, 67 p. [http://ific.auf.org/sites/default/files/@@Etude%20des%20portails-francophone-V02.pdf]

La formation sur les TIC/E a toute sa place dans les projets numériques de l'AUF tant par la formation des formateurs que par l'offre de formation diplômante en ligne en partenariat avec les universités francophones.

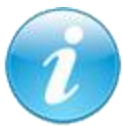 C'est dans une articulation à deux dimensions qu'une stratégie d'action (feuille de route) autour du numérique éducatif dans la région Asie-Pacifique pourrait atteindre ses objectifs ; deux faces évoluant en parallèle pour générer une dynamique autour et par des CNF et CNFp et la répandre ensuite dans l'entourage professionnel et universitaire de la région.

La formation sur les TIC/E en Asie pacifique peut s'inscrire de façon transversale dans les quatre missions du BAP et plus prioritairement dans la mission A : « Enseignement supérieur et transferts d'expertise » et la mission C « Numérique éducatif et services à la communauté universitaire francophone ». La première vise l'accompagnement des différents établissements membres de la région qui en font la demande dans leur volonté de mettre en place des projets interuniversitaires de formation francophone de niveau licence, master ou doctorat. La deuxième prévoit une orientation plus professionnalisante des formations de formateurs pour l'amélioration des compétences des enseignants en exercice et pour des futurs diplômés en quête de compétences numériques pour l'insertion professionnelle. Les missions B « Recherche et thématiques intégrées » et D : « Programmation scientifique et vie associative » auront à un moment ou un autre besoin de formation pour le renforcement de capacités d'acteurs concernés.

Dans ce contexte de synergie francophone autour du numérique éducatif, la formation sur les TIC/E devrait être développée dans les deux orientations suivantes :

1. renforcement des capacités de la communauté universitaire francophone à travers des sessions de formation à caractère professionnel organisées dans le réseau des Campus numériques francophones (CNF) et Campus numériques francophones partenaires (CNFp). L'essentiel de ces formations est accompli dans le cadre des attributions de la mission C : « Numérique éducatif et services à la communauté universitaire francophone » ;

2. accompagnement des différents établissements membres de la région dans leur volonté de mettre en place des projets interuniversitaires de formation francophone de niveau licence et de niveau master. L'essentiel de ces formations est accompli dans le cadre des attributions de la mission A : « Enseignement supérieur et transferts d'expertise ». Cette mission, si elle prend une nouvelle dimension avec la restructuration de l'AUF de sa politique d'accompagnement (plutôt que de substitution), est pourtant ancienne et prolifique[13].

## 2.1. État des lieux dans les pays CLV

La formation sur les compétences numériques a été l'un des maillons faibles constatés dans le paysage universitaire des pays visités. Nous entendons par formation, le renforcement des capacités numériques des enseignants et/ou des apprenants pour en faire des acteurs dans tout programme ou projet de réforme universitaire ou d'insertion professionnelle par le numérique. Pour des raisons multiples, notamment celles associées au facteur de la disponibilité, de la compensation, du contrôle de qualité, de la valorisation des acquis d'expériences, etc., la formation des formateurs sur les thèmes du numérique éducatif a été l'une des défaillances du système numérique dans plusieurs universités visitées. À part quelques initiatives de formation inscrites dans des projets de coopération comme l'ACU, les enseignants n'ont pas généralement de cadre institutionnel assurant une régularité de formation et de mise à niveau de leurs compétences numériques. Dans certains cas, c'est également dû à une inertie et une abstention à réclamer ce genre de mise à niveau. Cette inertie risque toutefois de se trouver bousculée par la réforme de l'enseignement supérieur programmée par l'ASEAN pour ouvrir ses frontières en 2015 avec une volonté d'uniformiser les formations et les diplômes à partir de 2017 sur le modèle LMD. Dès lors, des besoins en formation, en certification, en évaluation et en assurance qualité seraient inéluctablement nécessaires pour la mobilité des étudiants et des formateurs d'une institution à l'autre et d'un pays à l'autre.

---

[13] Voir la liste des formations francophones soutenues par l'AUF en Asie-pacifique. [https://www.auf.org/media/filer_public/bc/3d/bc3dcecc-e384-4b8f-9be5-dc42f185ab0c/liste_des_formations_francophones_en_ap_eng.pdf]

## 2.2. Les actions du BAP pour la formation sur les TIC/E

La région de l'Asie pacifique a été, comme toutes les autres régions de la Francophonie, concernée par les formations du type Transfer. Plusieurs ateliers et sessions de formation sur les TIC et les TICE ont été programmés et conduites dans les pays de la région notamment au Vietnam dans les nouveaux locaux du campus numérique francophone partenaire (CNFp).

Pendant la conférence de presse « Coopération universitaire francophone au Vietnam », organisée le 11 avril 2016 dans les locaux du CNFp de Hanoi, à l'occasion de la visite au Vietnam de Jean-Paul de Gaudemar, recteur de l'AUF, Sophie Goedefroit, directrice du Bureau Asie-Pacifique (BAP) de l'AUF, a annoncé que ce campus numérique, inauguré en avril 2015, est le fruit d'un partenariat entre l'AUF et l'Académie des sciences sociales du Vietnam (ASSV). Les différentes actions qui y ont été menées « montrent la diversité des besoins, mais aussi celle des réalisations », a-t-elle précisé. « En termes de formation de formateurs, nous avons, il y un peu moins d'un an, donné une présentation des nouvelles formations à distance pour les enseignants de français, a-t-elle poursuivi. En termes de réflexion, nous avons eu des ateliers avec des experts définissant les démarches de qualité au niveau des universités ». Des ateliers de formation ont été conduits sur la maîtrise de recherche et d'accès à la documentation scientifique, sur les techniques et les modules d'insertion professionnelle à destination des étudiants et des enseignants universitaires[14].

Cette nouvelle dynamique de formation conduite par les CNF et CNFp de la région gagnerait par contre à s'aligner sur les compétences définies dans le nouveau référentiel TIC/E de l'AUF.

## 2.3. Le référentiel de compétences TIC/E de l'AUF

Dans le domaine de la formation, l'AUF dispose de suffisamment de ressources et d'expériences pour proposer des pans entiers de coopération, voire un leadership dans l'accompagnement de projets et de programmes de formation (formations de formateurs ou formations diplômantes, en présentiel ou à distance, hybrides ou ubiquitaires).

---

[14] Les universités francophones cherchent à améliorer leur qualité. Le Courrier du Vietnam, 12/04/2016. [http://lecourrier.vn/les-universites-francophones-cherchent-a-ameliorer-leur-qualite/255250.html]

Depuis 2015, l'AUF comme d'autres porteurs d'idées sur l'innovation par le numérique, s'est prémunie d'un nouveau référentiel de compétences TIC/E[15] conçu à partir d'une vision de transversalité des compétences comme valeur favorisant la mobilité, le travail collectif, la prise en compte de la diversité par l'interopérabilité, ainsi que le libre accès à la connaissance. Cette préoccupation vise particulièrement à valoriser l'expérience et à diffuser les savoirs et les savoir-faire pour nourrir une meilleure gouvernance universitaire par les TIC et les TICE.

L'engagement de l'AUF sur la voie de l'interopérabilité des TIC/E par les référentiels de compétences provient de sa conviction que la FOAD est un vecteur important d'accès à l'enseignement supérieur pour un large public francophone qui ne peut y avoir autrement accès. L'AUF pourrait ainsi aider à sédentariser les talents et contribuer à lutter contre l'exclusion et la fuite des cerveaux.

Or, la réalisation de tels objectifs ne peut s'accomplir que par la mise au point d'un cadre commun de formation et de certification qui serait élaboré et adopté par ses propres établissements membres. Dans ce cas, un référentiel commun de formation servirait de catalyseur pour une harmonisation plus étendue des pratiques, une interopérabilité plus productive des ressources et une qualité plus valorisante des services. Un référentiel pédagogique viendrait ainsi multiplier la surface de la coopération académique et scientifique en offrant la possibilité d'un extraordinaire élargissement de l'offre de formation francophone en ligne : formations ouvertes et à distances et ressources éducatives libres, des cours en ligne ouverts et massif (CLOMs) francophones, etc. L'AUF pourrait aussi exporter ses compétences et son savoir-faire vers d'autres systèmes académiques régionaux et internationaux.

Une étude exploratoire pour l'identification des besoins en formation TIC/E dans les pays francophones du sud a été menée par l'AUF en 2016 afin de proposer un état des lieux représentatif devant contribuer à l'élaboration d'une stratégie francophone pour la formation des formateurs dans le domaine du numérique éducatif pour la francophonie universitaire, notamment dans les pays en développement et pour les universités en développement[16]. Elle pourrait fournir des indicateurs de choix pour une politique de formation de formateurs comme prévue par l'AUF à partir de son nouveau référentiel de compétences..

L'original dans le nouveau Référentiel TIC/E de l'AUF est sa modularité et sa flexibilité combinatoire pour constituer des formations clés en main ou adaptables selon les situations et les contextes des commanditaires. Elles peuvent prendre forme d'ateliers ou de parcours de formation. Un atelier pouvant cibler plusieurs compétences liées à une activité ou un projet ; un parcours se focalise généralement sur une seule compétence.

Ces formations, basées sur l'articulation de plusieurs compétences du numérique éducatif et leurs déclinaisons en savoirs, savoir-faire et savoir-être, peuvent servir à des fins pédagogiques d'exécution d'activités, de gestion de projets, d'administration de chantiers, de définition de parcours de formation ou de certification de profils métiers dans le domaine du numérique éducatif. Le potentiel combinatoire de ce référentiel est très large pour répondre à tous les besoins de formation sur le numérique éducatif.

---

[15] Cf. http://ific-auf.org/transfer/le-r%C3%A9f%C3%A9rentiel-tictice

[16] BEN HENDA Mokhtar. « Identification des besoins TIC/E dans les pays francophones du Sud ». AUF-IFIC, juin 2016 []http://ific.auf.org/sites/default/files/Rapport%20final%20TICE%20AUF2016.pdf

Depuis sa fondation en janvier 2015, le référentiel TIC/E de l'AUF a permis de concevoir des ateliers clés en main pour les parcours suivants :

| | |
|---|---|
| 1) Tutorat à distance<br>2) Piratage éthique<br>3) Ré-enchantement de la transmission des savoirs, savoir-faire, savoir-être au moyen des jeux vidéo à vocation pédagogique dans un contexte d'enseignement numérique<br>4) Fondamentaux de la scénarisation pédagogique<br>5) Jeu tangible LudiScen<br>6) Passage d'un enseignement conventionnel à l'apprentissage multimodal mobile (la classe inversée enrichie)<br>7) Risques cybernétiques<br>8) Organisation d'une classe/un cours inversé avec des outils pertinents et des applications en ligne adéquates<br>9) Création de document scientifique avec LaTeX<br>10) Évaluation dans les dispositifs d'apprentissage | 11) Parcours Déploiement d'une plateforme de « eformation » : installation, administration et intégration de contenu pédagogique - exemple et cas pratique : Moodle<br>12) Mise en place et gestion d'une communauté en ligne<br>13) Veille informationnelle<br>14) Développement des technologies éducatives<br>15) Indexation des ressources pédagogiques libres<br>16) Certification LPIC1<br>17) Installation technique et gestion informatique d'une PF de MOOC (Open EdX)<br>18) Scénariser un MOOC<br>19) Réaliser une vidéo pédagogique<br>20) Animer et piloter un MOOC<br>21) Intégration des ressources sur une PF de MOOC et gestion des fonctionnalités (Open EdX) |

Le nombre de ces ateliers et parcours de formation clés en main est censé augmenter au fil du temps, chaque fois qu'un nouveau besoin aurait été exprimé par un partenaire francophone. L'idée est de créer une banque de ressources de formation similaire au projet BASAR (BAnque de Scenarii d'Apprentissage Hybrides Réutilisables et Interopérables)[17].

## 2.4. Recommandations pour le BAP

Dans le cadre de la formation, le BAP est déjà sur une mission en cours « Mission A : enseignement supérieur et transferts d'expertise » qui vise à accompagner les différents établissements membres de la région qui en font la demande dans leur volonté de mettre en place des projets interuniversitaires de formation francophone de niveau licence et de niveau master. Cette mission vise également à mobiliser une offre d'expertise francophone à travers les pôles scientifiques régionaux dans le cadre d'une approche par projet.

Par cette mission, le BAP envisage conduire les actions suivantes :

- mettre en place une offre de formation pour les enseignants et producteurs de contenus centrée sur l'analyse, le référencement et la mutualisation de ces ressources ;
- élaborer puis développer une offre francophone d'expertise (approche par projet et démarche-qualité) pour l'accompagnement d'un projet de formation à distance ;
- élaborer puis développer une offre francophone d'expertise (approche par projet et démarche-qualité) pour l'accompagnement d'un projet de formation hybride (le bureau régional accompagne à ce jour plus de 40 formations présentielles). Des formations hybrides seraient très tributaires de parcours de formations sur les technologies numériques de l'enseignement à distance.

---

[17] Le Projet BASAR dont la mise en place est le résultat des efforts conjoints de quatre bureaux régionaux (BECO, BM, BMO et BEO) et du vice rectorat à la vie associative et la coordination des régions, a pour objectif principal la mise en place et l'alimentation d'une banque de scénarios hybrides destinée aux enseignants des Universités Francophones. Les Scénarios de cette banque concernent un maximum de domaines scientifiques ainsi que les trois niveaux de l'enseignement universitaire : Licence, Master et Doctorat [http://www.projetbasar.net/index.php/fr/]

À ces options de formation, le référentiel TIC/E propose des parcours clés en main. Le parcours « Indexation des ressources pédagogiques libres » concorde, par exemple, avec la dynamique actuelle autour du projet IDNEUF et du portail des REL francophones. À ce sujet, et dans la logique d'une approche par projet, des sessions de formation sur l'indexation des REL seraient fort recommandées, d'une part pour agir plus activement dans le projet du portail francophone des REL, mais aussi pour évoluer d'un premier niveau de conception de cours en ligne (formation déjà réalisée en nombre dans la région Asie pacifique) à un niveau avancé d'indexation et de référencement de ressources éducatives en ligne. Des documentalistes, bibliothécaires mais aussi des enseignants ayant suivi les formations sur la conception des cours en ligne, gagneraient en performance et en qualité de production s'ils sont impliqués dans des sessions de formations Transfer dans cette branche de compétences liées aux REL Ainsi les deux missions A et C créeraient un terrain de synergie pour former d'abord les enseignants universitaires sur le compétences TIC/E qu'il serait ensuite plus facile d'accompagner (dans le cadre de la mission A) pour assurer des formations hybrides dans leurs programme de formations de niveaux licence et master.

Différents ateliers ont été organisés en 2015 par le BAP en vue de structurer une offre francophone d'expertise dans le domaine de la démarche-qualité. C'est un objectif prioritaire que le BAP a défini pour des raisons de structuration puis de développement d'une culture de la qualité au sein des institutions d'enseignement supérieur et de recherche qui en font la demande. À travers l'élaboration de guides, il s'agit de mettre en place un ensemble accepté et partagé de références sur la démarche-qualité, mais aussi de proposer un pilotage de cette démarche. Ce qui explique que ces guides sont pensés comme des outils opérationnels/pédagogiques structurés qui offrent :

- Un cadre de référence qui vise à définir les objectifs de la démarche-qualité ainsi que les différentes étapes qui peuvent rythmer sa mise en œuvre ;
- Un guide pratique organisé en fiches qui regroupent l'ensemble des éléments ayant une influence sur la qualité des activités programmées. Chaque fiche se divise en plusieurs axes, chacun comportant une liste de questions ;
- Un dispositif de suivi et d'accompagnement opérationnel au service de l'analyse des résultats de la démarche-qualité et d'une restitution sous forme de recommandations.

Un équilibre a été recherché dans les orientations stratégiques proposées entre, d'une part, la création et le développement d'une culture de la qualité et, d'autre part, la spécificité des institutions membres de la Conférence générale des recteurs d'universités membres de l'AUF en Asie-Pacifique (CONFRASIE). L'objectif est d'accompagner les institutions d'enseignement supérieur et de recherche qui souhaitent développer leurs propres pilotages de la démarche-qualité, tout en contribuant à établir des références communes.

Pour la période 2016-2017, le BAP pourrait envisager, dans ce créneau de formation, de faire remonter au Référentiel TIC/E sa propre conception (et ressources associées) de parcours de formation sur la démarche-qualité. Ainsi il participerait activement à la banque de parcours de formation de l'IFIC.

# 3. RECHERCHE SCIENTIFIQUE AUTOUR DU NUMÉRIQUE ÉDUCATIF

La recherche dans le domaine du numérique éducatif est de nature pluridisciplinaire puisqu'elle fait appel aux sciences de l'éducation, à l'informatique, aux sciences de l'information et de la communication, à la psychologie cognitive, à la sociologie, aux sciences politiques, à l'économie du développement, etc. Elle s'applique à tous les domaines du savoir humain tant que ces savoirs sont concernés par l'usage des TIC et/ou transmis grâce aux TIC/E. Si ce croisement de disciplines est source de richesse, on constate dans les pays du Sud que la recherche dans ce domaine a du mal à se structurer, et que les chercheurs sont encore trop souvent isolés et peu ou mal reconnus par les institutions dans la progression de leurs carrières. L'une des principales raisons de ce déficit de valorisation, outre la fragmentation, est l'absence de publications scientifiques dans des revues à comités de lecture. Sans cette reconnaissance par les pairs, le travail des enseignants impliqués dans les TICE et la FOAD peine à être reconnu. La publication est nécessaire à la promotion dans une carrière et, plus largement, elle permet de faire connaître des travaux originaux, de souder une communauté scientifique, d'y organiser des débats et de faire avancer des idées.

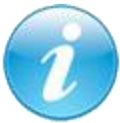 Ce n'est pas par hasard si certains des meilleurs chercheurs sont également d'excellents professeurs. Alors qu'il peut y avoir des chercheurs confirmés qui ne soient pas de bons enseignants, l'effort d'enseignement conduit souvent à de meilleurs résultats de recherche.

## 3.1. La situation de la recherche dans les pays CLV

Nous l'avons déjà signalé : la recherche dans les trois pays CLV est pratiquement déconnectée de l'activité pédagogique et de l'enseignement. Généralement, les politiques et les systèmes éducatifs dans les CLV ne sont pas définis dans une optique de priorité nationale attribuée à la recherche scientifique et aux technologies numériques. Chaque ministère met en œuvre des programmes de formation de ses propres techniciens et élabore des programmes de recherche à sa convenance et selon ses propres besoins. En outre, avec les hautes charges de l'enseignement et une formation insuffisante en recherche, aucune culture de recherche scientifique de haute qualité n'a pu se développer dans les universités des pays CLV. A ne citer que le cas du Cambodge où il y a un manque de professeurs nationaux de hauts niveaux qui pourraient enseigner les matières de technologies de pointe. Dans le cas de l'Université Royale de Phnom-Penh (RUPP), par exemple, seuls deux professeurs ont un diplôme de PhD sur un total de 119. Ce chiffre alarmant est un indicateur clé sur l'état de la recherche dans une université de prestige comme la RUPP, et par conséquent sur le statut de la recherche dans le pays en général.

Quant aux causes réelles de ce repli de la recherche scientifique, nous les avons déjà identifiées comme liées à des facteurs économiques, aux budgets insuffisants pour la recherche, à une carence de personnel enseignant qualifié, à la faiblesses de la production scientifique et la faiblesse des salaires, à la faiblesse des infrastructures technologiques et la précarité des conditions de travail dans les universités publiques, etc.

Néanmoins, conscients de ce décalage, les gouvernements du CLV insistent de plus en plus pour que la capacité de recherche soit développée dans les établissements d'enseignement supérieur, en particulier dans les universités qu'ils désignent prestigieuses et clés pour l'enseignement supérieur. Cette décision corrobore une tendance croissante dans le monde selon laquelle des activités de recherche de pointe sont concentrées dans un certain nombre d'institutions en vue d'y créer des pôles d'excellence avec des niveaux élevés de compétitivité.

### 3.2. L'appui francophone à la recherche dans la région Asie pacifique

L'appui à la recherche constitue déjà l'une des missions principales du BAP. Cette mission, rappelons-le, vise à accompagner le renforcement institutionnel en matière d'organisation et d'animation de la recherche et de la formation à la recherche, en privilégiant les capacités collectives de recherche et de formation pour la recherche au niveau d'équipes ou de laboratoires. Elle vise également à favoriser la constitution de réseaux de recherche collaboratifs pluridisciplinaires sur des thématiques intégrées portés par des membres de l'Agence et leurs partenaires de la région et hors région.

Dans le cadre de sa programmation 2014-2017, l'Agence universitaire de la Francophonie a organisé depuis 2014 une réflexion stratégique sur la contribution de la francophonie universitaire à de grandes thématiques pour proposer des actions favorisant la production et la diffusion des contributions francophones. Un comité ad hoc s'est réuni, à Paris le, 7 octobre 2014, à l'initiative de l'AUF, pour produire une liste de thématiques et de problématiques autour desquelles l'AUF souhaite mobiliser ses établissements membres. Les conclusions de ce comité ont été amendées puis approuvées par les instances de décembre 2014. Quatre appels à projets ont été ouverts par le Bureau Asie-Pacifique adressés à tous les établissements et institutions membres de l'AUF dans la région en vue de créer une synergie de renforcement du potentiel de recherche et de formation novatrices dans les domaines de l'eau et gestion des ressources naturelles, de l' Énergies renouvelables, de la Gestion des crises et des conflits et de la Santé publique. Le BAP a mis en place un dispositif d'accompagnement des projets interuniversitaires de recherche en vue d'assurer le suivi opérationnel au service du pilotage des activités à développer au sein des projets.

Dans son intervention à l'occasion du Séminaire régional de recherche francophone 2015 « Évolution de la recherche en Asie du Sud-Est et les nouvelles pistes », Madame Sophie Goedefroit, Directrice du BAP de l'AUF a mis l'accent sur les actions que le bureau régional a réalisées ou soutenues dans le domaine de la recherche, notamment le soutien à la formation doctorale, le soutien aux rencontres scientifiques et aux publications scientifiques francophones, la réflexion stratégique sur la contribution de la francophonie universitaire aux grandes thématiques intégrées. « *Le soutien de l'AUF couvre toutes les disciplines (l'environnement, la santé, le patrimoine, les sciences politiques...) et l'ensemble des pays de la région : de l'Asie continentale jusqu'au Japon, le Vanuatu et la Nouvelle-Calédonie, avec des ramifications dans les autres continents de la francophonie universitaire.* », a rajouté Madame Goedefroit.

Les initiatives de recherche menées par le BAP, gagneraient eux-aussi de s'inscrire dans la politique générale de recherche de l'AUF. Dans le domaine du numérique éducatif, un rapprochement avec le nouveau réseau de recherche francophone en éducation numérique (AREN)[18] serait très utile pour les projets de recherche francophone en Asie pacifique.

---

[18] AREN : un méta-portail pour faciliter l'accès aux ressources universitaires francophones [http://aren-portail.org/]

### 3.3. AREN, le réseau de recherche francophone en éducation numérique

La recherche en éducation numérique se caractérise par un éclatement thématique qui irrigue des disciplines scientifiques différentes. Une recherche de qualité en TIC/E se doit d'être un vecteur de transversalité et de transdisciplinarité. Or, dans les pays en voie de développement, le domaine se caractérise généralement par l'émergence régulière d'effets de mode tout en reposant sur un socle pédagogique qui évolue très peu.

En Francophonie, beaucoup de compétences ont été formées en éducation numérique grâce à des masters spécialisés en ingénierie de formation et par l'intermédiaire d'ateliers de formation organisés par l'AUF. Cependant les enseignants qui les ont suivis sont souvent isolés au sein de leurs structures. Par ailleurs le nombre de docteurs du domaine est faible, le nombre d'encadreurs disponibles est lui aussi insuffisant.

Intimement liée à la mission de formation des universités, l'activité d'exploration, de recherche, de production et de partage des savoirs est plus que jamais ancrée dans la mission de l'Agence universitaire de la Francophonie. Pour la période 2014-2017, cette mission s'articule autour de deux objectifs, dont bénéficient en priorité les universités en sortie de crise ou en développement :

1. favoriser le développement d'équipes de recherches internationales engagées dans des problématiques émergentes. Les formes de soutien consistent à encourager la création d'équipes de recherche francophones de dimension internationale, le développement de projets régionaux et inter-régionaux, et la coopération entre équipes en émergence.

2. favoriser l'innovation et le renforcement de la pertinence sociale, scientifique et technologique des résultats de recherche. Les formes de soutien comprennent l'organisation de séminaires de rédaction de projets et d'articles scientifiques, le soutien à l'édition électronique, l'aide aux réseaux universitaires, notamment ceux de jeunes chercheurs.

Consciente de cette réalité, la communauté francophone de la recherche en TICE s'est dotée d'un cadre structurel fédérateur capable de mettre en synergie toutes ces potentialités souvent éparpillées et déconnectées entre-elles dans l'espace et dans le temps. AREN se propose ainsi d'assumer le rôle de structure fédérative de la recherche francophone en TICE dotée d'outils, de moyens et d'une stratégie d'action prospective pour l'innovation et la création de l'excellence.

L'AUF a déjà participé à la création de plusieurs réseaux de recherche dans le domaine du numérique éducatif dont Res@Tice (2005), TICER (2007) et le réseau « MIRRTICE » (2013). Le domaine des TICE continue encore à être l'un des deux champs prioritaires qu'a choisi l'AUF dans sa programmation quadriennale 2014-17. Dans ce cadre, et par l'intermédiaire de l'IFIC, l'AUF s'est proposée depuis la fin 2015 de relancer un réseau de chercheurs autour de la thématique des TICE avec pour objectif direct et projet constitutif, l'appui à l'émergence de jeunes scientifiques francophones. L'idée a suscité la création de l'Alliance pour une recherche en éducation numérique (AREN) afin de dynamiser la recherche en éducation numérique dans les pays francophones émergents. Les débats constitutifs de la création d'AREN ont permis de dégager trois priorités d'actions :

1. Appuyer la publication scientifique pour aider les doctorants et les enseignants-chercheurs à progresser dans leur carrière ;
2. Créer des mécanismes de socialisation pour favoriser les échanges scientifiques ;

3. Favoriser les partenariats interuniversitaires notamment pour l'encadrement de doctorants dans le domaine des TICE.

L'AUF s'engage aussi à inscrire ces priorités dans la programmation de l'IFIC, à les traduire en activités et actions définies en concertation avec les signataires de la charte de création du réseau.

Parmi les premières actions de ce jeune réseau, AREN a lancé en janvier 2016 un appel à propositions intitulé « coup de pouce AREN 2016 » doté de 55 000 euros. L'objectif général de l'appel à propositions est d'apporter un appui à la publication scientifique, dans des revues imprimées (papier) ou numériques, dans les disciplines liées au numérique éducatif, à la formation à distance, aux usages des technologies innovantes pour l'enseignement. Les porteurs des projets retenus recevront une aide financière et un accompagnement méthodologique en vue de soumettre un article scientifique à une revue francophone à comité de lecture spécialisée dans ces champs disciplinaires. Il s'agit d'aider les enseignants et enseignants-chercheurs à progresser dans leur carrière à l'aide de la publication scientifique des résultats de leurs travaux.

## 3.4. Recommandations pour le BAP

L'activité de recherche est inscrite pour le BAP dans la « mission B : « Recherche et thématiques intégrées ». Cette mission vise à accompagner le renforcement institutionnel en matière d'organisation et d'animation de la formation à la recherche, à travers la mise en réseau d'écoles doctorales dans la région et au-delà, et ceci en privilégiant les capacités collectives d'équipes ou de laboratoires. Cette mission vise également à favoriser la constitution de réseaux de recherche collaboratifs pluridisciplinaires portant sur des thématiques jugées prioritaires dans la région et portés par des membres de l'Agence.

Dans une récente étude effectuée par l'auteur de ce rapport au compte de l'AUF en 2016[19], il a été question d'examiner les propositions des enseignants-chercheurs dans les universités francophones du Sud, pour appuyer la recherche autour des TIC/E dans leurs institutions. Les réponses ont été très diverses dont les plus significatives ont été concentrées autours des équipements adéquats, de formations à tous les acteurs universitaires (étudiants, enseignants et formateurs professionnels), puis un accès internet à très haut débit. Plusieurs autres critères ont été signalés, mais malgré leur importance, ces critères n'avaient pas obtenu un large consensus. Parmi ces critères, signalons notamment « du Wifi partout », « des technologies mobiles », « un personnel qualifié », « des réseaux interinstitutionnels de recherche », « la reconnaissance des diplômes TICE », « plus de temps libre pour la recherche », etc.

Dans les conditions propres à l'état de la recherche dans les pays de l'Asie pacifique, le BAP pourrait entreprendre des mesures d'appui à la recherche autour des quatre axes suivants :

1) Appuyer la publication scientifique pour aider les doctorants et les enseignants-chercheurs à progresser dans leur carrière ;

---

[19] 21.    BEN HENDA Mokhtar. « Identification des besoins TIC/E dans les pays francophones du Sud ». AUF-IFIC, juin 2016 [http://ific.auf.org/sites/default/files/Rapport%20final%20TICE%20AUF2016.pdf]

2) Créer des mécanismes de socialisation pour favoriser les échanges scientifiques ;

3) Favoriser les partenariats interuniversitaires notamment pour l'encadrement de doctorants dans le domaine des TICE.

Plusieurs suggestions peuvent émerger sous ces axes clés :

- Prévoir des sessions de sensibilisation orientées sur la culture de la recherche scientifique notamment sur les handicaps culturels et psycho-cognitifs de la recherche : la rétention de l'information (par individualisme), le plagia (par insouciance ou paresse intellectuelle) et les droits d'auteurs (par ignorance ou insouciance) ;

- Prévoir des formations pour les chercheurs sur les techniques et les modalités de la rédaction scientifique et de l'édition numérique ;

- Appuyer l'usage des réseaux sociaux académiques comme vecteur de promotion, de mutualisation et de canevas d'échange et de diffusion des publications scientifiques entre les universitaires …

La principale démarche du BAP pour l'appui à la recherche viendra sans doute dans le cadre de la mission B : « Recherche et thématiques intégrées », notamment au niveau de l'accompagnement des formations francophones de niveau doctoral.

# FEUILLE DE ROUTE

Ce dernier point de l'expertise est en réalité prévu pour être une étape de co-construction d'une feuille de route 2016-2017 avec les équipes de l'AUF en Asie pacifique (BAP, CNF et CNFp) incluant un programme de formation de formateurs à deux dimensions francophones, l'une locale propre à la région de l'Asie-Pacifique et l'autre internationale conforme à la politique numérique générale de l'AUF.

La feuille de route proposerait donc des paramètres et des indicateurs de définition et d'expérimentation de dispositifs pédagogiques francophones de formation dans une politique cohérente de qualité et de compétitivité non seulement pour les deux dernières années de la programmation quadriennale en cours (2014-2017), mais aussi pour les années d'après. Elle devrait donner des indicateurs sur les contenus de formation, les modèles pédagogiques et les publics cibles.

En somme, la feuille de route tiendrait compte des critères suivants :

- *Définir un contenu de formation qui répondrait aux besoins locaux et régionaux déduits de l'enquête d'expertise* : l'enquête a décrit l'état de la situation dans les trois pays CLV en termes de compétences TIC/E, de moyens et de ressources pédagogiques, de politiques nationales relatives aux TIC/E, de besoins et de nécessités dont les communautés universitaires ont besoin. La feuille de route devrait en tenir compte pour satisfaire aux besoins des différents acteurs concernés ;

- *Définir un contenu de formation qui anticipe les besoins futurs de la région conformément aux programmes du numérique éducatif francophone* : la stratégie de l'AUF, comme définie dans son modèle de gouvernance par le numérique éducatif, est surtout de veiller à garder un haut niveau de qualité et de compétitivité avec les référentiels internationaux. En Asie pacifique, les structures anglophones de l'ASEAN, notamment l'ACU, constituent un challenger à « affronter » avec de la qualité et de la compétitivité. Il s'agit aussi pour l'AUF de créer en interne une forte synergie entre les différentes régions de la Francophonie pour l'échange et la collaboration. Toutes les régions, et donc tous les bureaux régionaux, sont sollicités à participer activement aux grands projets du numérique éducatif francophone. La région de l'Asie-Pacifique devrait à ce titre disposer de compétences locales capables de répondre aux attentes du numérique francophone international. Les formations Transfer et le nouveau référentiel des compétences TIC/E développé par l'IFIC font désormais office de locomotive pour pousser les régions vers l'acquisition de compétences génératrices d'innovation.

- *Cibler un public étudiant en formation universitaire* : la feuille de route devrait nécessairement tenir compte de la réforme entreprise par le BAP dans sa nouvelle politique d'accompagnement de formation universitaire francophone au niveau du LMD afin de répondre à sa première mission (parmi 4)[20] : « Enseignement supérieur et transferts d'expertise ». Rappelons ici que le Consortium d'appui aux formations francophones, réuni en octobre 2014 à Hô Chi Minh-Ville, a annoncé que la programmation quadriennale 2014-2017 et la politique scientifique sur laquelle elle s'appuie impliquent le passage d'une logique de soutien ou de substitution à une logique d'accompagnement par projets interuniversitaires comportant la définition d'objectifs ciblés et d'indicateurs mesurables. La mise en œuvre de ces nouvelles modalités d'accompagnement a amené la Commission régionale des experts (CRE) à arrêter certains soutiens (logique de substitution). La Francophonie a ainsi changé sa politique d'appui à l'usage du français en ne finançant plus entièrement les formations en langue française. Elle prévoit plutôt intervenir en appui à des structures universitaires dans le déroulement de leurs programmes LMD. Les détails de cette stratégie sont décrits dans 4 guides préparés par le BAP pour l'accompagnement de cycle de formation universitaire francophone au niveau licence[21], master[22] et doctorat[23] avec un guide de démarche qualité pour les formations de licence et de master[24]. Par ces outils, le BAP a souhaité mettre à disposition des acteurs de la francophonie universitaire de la région plusieurs outils afin qu'ils puissent devenir eux-mêmes les experts dans la conduite de leur projet de formation.

---

[20] Les trois autres missions sont : « Recherche et thématiques intégrées », « Numérique éducatif et services à la communauté universitaire francophone », et « programmation scientifique et vie associative ».

- *Cibler un public francophone en formation professionnelle* : l'AUF est aussi un acteur francophone ouvert sur le monde professionnel par la formation et la recherche. Elle prévoit dans ce sens des formations à vocation professionnalisante pour les futurs diplômés. Dans le but de favoriser l'employabilité et l'insertion professionnelle des jeunes francophones, la feuille de route devrait prévoir des formations du type Transfer sur des compétences innovantes du numérique éducatif. Le BAP a déjà introduit, de mars à avril 2016, une expérimentation de formation en mode hybride du cours « Module d'insertion professionnelle »[25]. Des formations professionnalisantes pour le renforcement des compétences numériques menées au sein des CNF et CNFp au profit de futurs professionnels francophones serait d'un grand appui à l'insertion professionnelle et d'un soutien à la langue française dans le milieu professionnel ;

- *Opter pour des modèles pédagogiques présentiel, à distance et hybrides selon la nature, la durée et le public cible concerné par chaque formation* : une formation sur les TIC/E, dans les conditions décrites ci-avant pourrait être montée selon plusieurs modèles pédagogiques. L'AUF participe à et dispense plusieurs types de formation : ponctuelles et/ou périodiques, courtes et/ou de durée moyenne ou longue. Une formation ponctuelle et de courte durée (de 2 à 3 jours) serait par exemple en mode présentiel destinée à la sensibilisation de décideurs, chefs d'établissements… Elle peut être aussi périodique (chaque moi, trimestre ou année). Une formation ponctuelle ou périodique peut être aussi de durée moyenne (5 à 7 jours) très pratique en mode présentiel pour la formation des enseignants et formateurs. Une formation de longue durée peut être menée à distance ou en mode hybride dans un cycle de formation universitaire (semestriel ou annuel). Plusieurs critères entrent en compte pour déterminer le modèle pédagogique d'une formation.

---

[21] GUIDE. Un dispositif régional au sein d'un réseau associatif mondial. Accompagnement d'un cycle de formation universitaire francophone - Niveau licence. https://www.auf.org/media/filer_public/89/9c/899c71ff-c8f6-41cf-998a-6a56fe9dbf50/guide_accompagnement_licence.pdf

[22] GUIDE. Un dispositif régional au sein d'un réseau associatif mondial. Accompagnement d'un cycle de formation universitaire francophone – Niveau Master. https://www.auf.org/media/filer_public/0f/b7/0fb791df-7ee0-441c-a8f4-bda70ca0f7da/guide_accompagnement_master.pdf

[23] GUIDE. Un dispositif régional au sein d'un réseau associatif mondial. Accompagnement d'un cycle de formation universitaire francophone – Niveau Doctorat. https://www.auf.org/media/filer_public/e1/45/e1456f87-1b97-49ce-bb94-e5492f4ac12c/guide_accompagnement_doctorat.pdf

[24] GUIDE. Démarche qualité : formation de licence, formation de master.

[25] Formation à l'insertion professionnelle via un dispositif hybride. https://www.auf.org/actualites/formation-linsertion-pro-hybride/

La feuille de route donnera un maximum d'éléments de description à des projets de formation selon les critères ci-haut mentionnés.

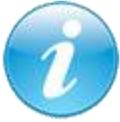

Cette feuille de route est une proposition formulée dans le cadre de l'expertise en cours. Elle fournira des éléments de programmation au BAP qui en retiendra ce qu'il jugera conforme avec la stratégie régionale pour le numérique éducatif. Elle viendra en complément à un programme d'activité déjà prévu par le BAP pour 2015 et 2016.

## 1. ACTIVITÉS DÉJÀ PROGRAMMÉES PAR LE BAP POUR 2015-2016

Une bonne partie de la programmation 2015 a déjà été réalisée. Il reste donc à valider dans la feuille de route 2016-2017 les activités déjà retenues dans la programmation du BAP pour l'année en cours (2016) et prévoir des activités nouvelles et innovantes pour 2017.

Les activités suivantes ont été décidées par le BAP comme prioritaires pour 2016 suite à un sondage réalisé en juillet 2015 par le bureau de la Conférence régionale des recteurs des établissements membres de l'AUF en Asie-Pacifique (CONFRASIE), en étroite concertation avec le bureau régional de l'Agence universitaire de la Francophonie pour identifier les besoins prioritaires de ses institutions membres en tenant compte des systèmes universitaires nationaux qui y sont représentés, et de la diversité qui les caractérise. Il s'agit surtout de définir des orientations stratégiques à mettre en œuvre dans le cadre d'une offre francophone d'expertise centrée sur la démarche-qualité :

### 1.1. Rédaction d'un guide démarche-qualité

Dans le cadre de la programmation quadriennale 2014-2017 du BAP, l'offre francophone d'expertise centrée sur la démarche-qualité est pensée comme un processus de renforcement continu des institutions et des compétences. Cette offre vise à structurer puis développer une culture de la qualité au sein des institutions qui en font la demande. La 11$^e$ session de la CONFRASIE a été l'occasion de définir des orientations stratégiques à mettre en œuvre dans ce cadre. Il s'agit de mettre en place un ensemble accepté et partagé de références sur la démarche-qualité, mais aussi d'en proposer un pilotage.

Compte tenu des résultats d'une enquête réalisée par le BAP, l'année 2016 est consacrée à l'élaboration de ces référentiels de démarche-qualité pour la définition et la mise en place d'un « plan de développement d'une institution ». Ce guide (outil) doit tenir compte des deux processus qui caractérisent la dynamique universitaire régionale : une volonté d'autonomisation des établissements et une recherche de reconnaissance tant régionale qu'internationale des projets qu'ils développent. L'élaboration de ce guide s'inscrit donc dans un contexte universitaire en pleine mutation, dont les perspectives développées par l'ASEAN en relation avec l'enseignement supérieur et la recherche constituent le troisième processus.

Jusqu'à cette date, plusieurs réunions ont été tenues à Hanoï, Danang et Ho Chi Minh Ville pour préparer ces guides destinés très particulièrement au programme d'accompagnement de cycles de formations universitaires francophones au sein des universités partenaires. Deux autres ateliers ont été programmés :

a) Un premier atelier réalisé pendant la semaine du 9 mai 2016 à Hanoï (Vietnam) pour produire un projet de guide qui se structure autour de quatre parties suivantes :

   1. Un cadre de référence qui vise à définir les objectifs de la démarche-qualité ainsi que les différentes étapes qui peuvent rythmer sa mise en œuvre ;
   2. Un guide pratique organisé en fiches qui regroupent l'ensemble des éléments jugés essentiels à la définition et la mise en place d'un plan de développement d'une institution ;
   3. Un dispositif de suivi et d'accompagnement opérationnel au service de l'analyse des résultats de la démarche-qualité et d'une restitution sous forme de recommandations ;
   4. Un glossaire qui vient préciser les différents concepts et expressions utilisés dans le cadre de ce document.

Ce premier atelier s'est concentré sur la partie « guide pratique ». Ce qui suppose de recueillir, traiter et restituer préalablement les informations jugées essentielles à la définition et la mise en place d'un plan de développement d'une institution, et de proposer aux participants à l'atelier un projet de guide structuré en axes - composantes - questions.

b) Un second atelier est prévu pour la fin du mois de septembre 2016 à Phnom Penh. Il s'agit de finaliser ce qui aura été réalisé au cours du premier atelier (glossaire), mais surtout d'inscrire certains axes et composantes dans une logique prospective (méthode par scénarios). C'est tout l'enjeu de ce second atelier. Comme pour tous les guides actuellement produits par le bureau de la CONFRASIE, ce guide sera finalisé début 2017 suite à des tests auprès des institutions membres de la CONFRASIE qui se sont portées volontaires en vue de tester cet outil.

Or, ce projet de rédaction de guides destinés à l'accompagnement de cycles de formation LMD dans les universités francophones de l'Asie pacifique s'inscrit plutôt dans la mission A et touche moins les autres missions, notamment la mission C du « numérique éducatif » dans laquelle des ateliers de formation de formateurs sur les TIC/E sont essentiellement prévus.

## 1.2. Ateliers de formation sur les TIC/E

Sur le plan de la formation des formateurs, le BAP a continué à offrir des formations de formateurs sur le modèle des ateliers Transfer[26]. Une série de formation sur les TIC/E a marqué les plans d'activités de plusieurs CNF et CNFp de la région Asie pacifique. L'un des derniers qu'on pourrait citer à titre d'exemple est celui organisé du 29 au 31 août 2016 par le Campus numérique francophone de Hô Chi Minh-Ville (CNF HCMV) intitulée : « Évaluation de cours par les étudiants : Approches, outils et adaptation des modèles existants dans le contexte asiatique ».

Cette formation s'est adressée uniquement aux stagiaires résidents de la région Asie-Pacifique, aux enseignants-chercheurs ou doctorants en sciences de l'éducation ou sciences humaines et sociales, aux chercheurs ou cadres des services d'assurance qualité des universités, aux enseignants-chercheurs des disciplines scientifiques intéressés par la recherche en éducation. Elle est le prototype des formations classiques programmées par les CNF et CNFp depuis longtemps. Or, avec le nouveau référentiel des compétences de l'AUF et les dernières études francophones qui ont identifié les besoins de formation dans les différentes régions de la Francophonie, les ateliers de formation des CNF et CNFp de la région Asie pacifique gagneraient à être alignés sur une stratégie numérique régionale inspirée des orientations stratégiques de l'AUF pour le numérique éducatif et utilisant des référentiels et des ressources adaptées.

---
[26] Ateliers Transfer. [http://www.transfer-tic.org/rubrique3.html]

Dans les points suivants, il sera question de définir les activités relevant de la mission C « Numérique éducatif et services à la communauté universitaire francophone » au titre des deux dernières années du quadriennal 2014-2017.

## 2. ACTIVITÉS PROPOSÉES POUR LA FEUILLE DE ROUTE 2016-2017

Les objectifs généraux suivants sont proposés pour figurer dans la programmation quadriennale du BAP pour 2016-2017 au titre de la mission C « Numérique éducatif » :

- Renforcer le vivier d'experts TIC/E à travers des formations de formateurs régionales, en s'appuyant notamment sur le nouveau référentiel de compétences TIC/E de l'AUF ;
- Diversifier les domaines de formations, leurs modèles pédagogiques et leurs publics cibles pour créer un contexte cohérent, favorable à l'innovation et l'insertion professionnelle :
- Créer des offres de formation hybrides liées à la mission « Enseignement supérieur et transferts d'expertise » pour l'accompagnement de cycles de formation universitaires francophones en LMD ;
- Créer un campus de référence (i.e. le campus numérique de Hanoï), à partir duquel sera structuré le réseau de formations de formateurs TIC/E, à échéance fin 2016 ;

Aussi, et en prenant compte de l'articulation opérationnelle entre les trois axes annoncés en début des recommandations, la feuille de route sera composée de trois principaux volets : développement de ressources numériques éducatives, renforcement des capacités autour du numérique éducatif et appui à la recherche dans le domaine du numérique éducatif.

### 2.1. Des activités relative au développement des ressources numériques éducatives, leur gestion, organisation et distribution

Les ressources numériques éducatives sont à la base de toute stratégie de numérique éducatif et de programme de formation sur les TIC/E. C'est une matière première qui existe à la source dans toutes les structures universitaires et de formation. Il est donc fondamental de commencer par rassembler les ressources existantes et de les restructurer pour les intégrer dans un système pédagogique cohérente.

Il faudrait aussi prévoir une stratégie d'action pour systématiser la production, collecte et le traitement des nouvelles ressources en vue de les intégrer systématiquement dans un processus de création de réservoir d'objets pédagogiques interopérable. Cela implique à la fois la récupération de ressources existantes et la production de nouvelles par les acteurs francophones partenaires.

Les activités suivantes peuvent être proposées dans la feuille de route 2016-2017

### 2.1.1. Ressources pédagogiques : Étude de l'existant

| | |
|---|---|
| *Type d'activité* | **Faire une étude de l'existant pour inventorier les acteurs institutionnels producteurs potentiels de ressources éducatives numériques dans la région Asie pacifique** |
| *Argumentaire* | L'une des lacunes observées pendant l'enquête de terrain menée dans la région Asie Pacifique est l'absence de toute démarche de recensement et collecte des supports de cours pour l'échange et la mutualisation. Les raisons derrière cette lacune sont à la fois la réticence des enseignants à vouloir déposer leurs supports de cours et l'absence de compensation (pécuniaire ou scientifique) de la part des institutions et des Ministères qui les motiveraient le faire.<br><br>Dans l'attente de trouver la meilleure façon d'accéder à ces ressources mal exploitées, il faudrait déjà les identifier et les recenser pour se faire une idée sur leur nature, type, envergure, qualité, coûts, etc.<br><br>Cette opération nécessite une enquête de terrain pour identifier les acteurs producteurs de ressources et estimer l'ampleur de l'opération de collecte et traitement. |
| *Objectifs* | - Faire l'inventaire des institutions universitaires francophones ou partiellement francophones pouvant disposer de ressources pédagogiques numériques ou non ;<br>- Mener une étude exploratoire au sein de ces institutions partenaires pour identifier les structures qui gèrent les programmes et les rapports avec les enseignants pour identifier les parties (personnes/équipes) productrices de ressources pédagogiques |
| *Mission cadre* | Mission C « Numérique éducatif » en collaboration avec la mission A « Enseignement supérieur et transfert d'expertises » |
| *Prérequis* | - Avoir une vue globale sur le système éducatif des pays de la région ;<br>- Disposer d'un répertoire général des institutions universitaires dans chaque pays ;<br>- Avoir une connaissance de la typologie des cours universitaires ;<br>- Maîtriser un outil de gestion d'enquête pour systématiser la collecte des données ; |
| *Responsabilité* | Équipe d'enquêteurs au niveau des CNF et CNFp du BAP sous responsabilité d'un chargé de mission « Patrimoine pédagogique numérique » |
| *Public cible* | - Institutions universitaires publiques et privées ;<br>- Centres de formation ;<br>- Bibliothèques universitaires ;<br>- Bibliothèques de recherche ;<br>- Etc. |
| *Date de démarrage* | - Octobre 2016 |
| *Durée* | - 3 mois |
| *Produits escomptés* | - Catalogue/inventaire des institutions universitaires potentiellement détentrices de ressources pédagogiques numériques ou non ;<br>- Coordonnées des personnes ressources dans ces institutions qui seraient des interlocuteurs directs pour la suite de l'activité. |
| *Types de formations nécessaires* | - Former une (des) personne(s) ressource(s) dans les bibliothèques et centres de documentation des institutions partenaires pour assurer la collecte et l'indexation des ressources éducatives disponibles ;<br>- Former une (des) personne(s) ressource(s) sur les techniques de la conduites d'enquête |
| *Type(s) de compétences(s) associées (Référentiel AUF)* | - (C1073) Formuler une question de recherche sur la base de sa connaissance de la littérature du domaine et de sa pratique du terrain<br>- (C1029) Concevoir et mettre en œuvre un dispositif de recherche permettant de répondre à la question posée<br>- (C1113) Recueillir, traiter et analyser les données<br>- (C1023) Communiquer le résultat de ses recherches |
| *Coûts prévisionnel* | - Coûts à estimer selon les grilles AUF pour un chargé de mission de 3 mois « Patrimoine pédagogique numérique » entre les CNF et CNFp<br>- Coûts d'un parcours de formation de 3 jours pour l'équipe chargée de l'enquête |
| *Indicateur de qualité* | - Exhaustivité du recensement ;<br>- Qualité des données d'identification et de contact ;<br>- Taux d'engagement des partenaires régionaux (par pays) ;<br>- Qualité des rapports produits. |

### 2.1.2. Ressources pédagogiques : collecte et traitement

| | |
|---|---|
| *Type d'activité* | ***Passer des conventions et procéder à la collecte des ressources pédagogiques auprès des institutions concernées*** |
| *Argumentaire* | La collecte des ressources pédagogiques peut avoir des aspects relativement complexes selon les techniques utilisées et les modalités juridiques mises en œuvre.<br><br>Les difficultés techniques sont engendrées par la nature des ressources (numériques ou papier) et la forme matérielle des données récupérées (document intégral ou seules métadonnées).<br><br>Toutes les solutions sont envisageables techniquement à condition de pouvoir accéder à la ressources par une opération de recherche en ligne.<br><br>Les difficultés juridiques sont déterminées par les questions de droits d'auteurs et des modalités mises en place pour l'exploitation légale des ressources. Il est fortement recommandé dans le contexte AUF d'encourager les droits d'auteurs *Creative Commons* quitte à « acheter » les droits d'auteurs par négociation. Il est également conseillé dans ce cas de figure d'utiliser l'argumentaire de la valorisation et de la visibilité scientifique internationale des institutions et des enseignants dépositaires de ressources dans un réservoir de REL ouvert et interopérable.<br><br>(cf. suite argumentaire sous §1.4.) |
| *Objectifs* | - Proposer et signer des conventions de cession de droit avec les institutions et les enseignants pour leurs ressources pédagogiques ;<br>- Établir un inventaire typologique et quantitatif des ressources pédagogiques rassemblées (genre, volume, niveau et granularité de chaque type de ressource) ;<br>- Identifier tout autre type de supports pédagogiques francophones éparpillés et mal exploités dans les institutions universitaires francophones ;<br>- Envoyer au portail IDNEUF des REL francophones rassemblées par le BAP (ses propres ressources de formation ou es ressources de ses partenaires conventionnés) ; |
| *Mission cadre* | Mission C « Numérique éducatif » en collaboration avec la mission A « Enseignement supérieur et transfert d'expertises » |
| *Prérequis* | - Connaître les textes législatifs du fonctionnement des institutions universitaires ;<br>- Avoir une connaissance des différentes formes de licences et des droits d'auteurs ;<br>- Avoir une culture des licences libres et des *Creative commons* ; |
| *Responsabilité* | Équipe d'enquête des CNF et CNFp du BAP sous responsabilité d'un chargé de mission « Patrimoine pédagogique numérique » |
| *Public cible* | - Institutions universitaires publiques et privées ;<br>- Centres de formation ;<br>- Bibliothèques universitaires ;<br>- Bibliothèques de recherche ; |
| *Date de démarrage* | - Janvier 2017 |
| *Durée* | - 6 mois |
| *Produits escomptés* | - Catalogue/inventaire des ressources pédagogiques (numériques ou non) recensées et compilées ;<br>- Rapports sur l'état de la collection des ressources rassemblées (documents primaires ou métadonnées, texte et/ou multimédia, langues de supports, indexées ou non, en cours d'utilisation ou archivés, cours ou exercices, niveau de granularité, etc.). |
| *Types de formations nécessaires* | - Exploiter les formations proposées sous l'activité A<br>- Prévoir une formation sur l'indexation des ressources éducatives libre |
| *Type(s) de compétences(s) associées (Référentiel AUF)* | - (C1009) Analyser-indexer-référencer les ressources numériques : « Assurer le traitement intellectuel de documents, extraire des documents les données textuelles, conceptuelles ou factuelles, les codifier ou les formaliser en vue d'alimenter une banque de données ou d'élaborer un produit documentaire électronique »<br>- (C1037) Conduire un processus de production d'information : « Créer une information considérée comme un processus qui se fonde sur la compréhension du fait que le but, le message et la livraison de l'information sont des actes intentionnels de production. Reconnaissant la nature de la production d'information, les experts se tournent vers les processus sous-jacents ainsi que le produit final à évaluer pour étudier de façon critique l'utilité de l'information ». |
| *Coûts prévisionnel* | - Coûts à estimer selon les grilles AUF pour un chargé de mission de 6 mois supplémentaires « Patrimoine pédagogique numérique »<br>- Coûts d'un atelier de formation de 5 jours pour l'équipe chargée de l'indexation et du contrôle des ressources pédagogiques rassemblées. |
| *Indicateurs de qualité* | - Nombre de partenaires conventionnés ;<br>- Volume des ressources répertoriées ;<br>- Qualité des rapports produits |

### 2.1.3. Ressources pédagogiques : diffusion et communication

| | |
|---|---|
| *Type d'activité* | ***Développer un système d'accès ouvert et à distance aux ressources pédagogiques en ligne*** |
| *Argumentaire* | Un réservoir d'objets pédagogiques est ouvert dans les deux sens : recevoir de nouvelles ressources mais aussi communiquer les ressources existantes.<br><br>Le BAP pourrait exploiter en amont les ressources rassemblées à son niveau pour en faire un entrepôt régional de ressources pédagogiques. Si cette opération n'est pas dans ses moyens actuels, il pourrait se contenter de déposer ses ressources dans le portail francophones des REL avec identification de leurs sources pour pouvoir les interfacer ensuite sur ses propres systèmes d'information.<br><br>Le portail IDNEUF actuel ([www.idneuf.org](www.idneuf.org)) permet de faire des recherches par pays mais aucun pays de l'Asie pacifique n'est encore répertorié !<br><br>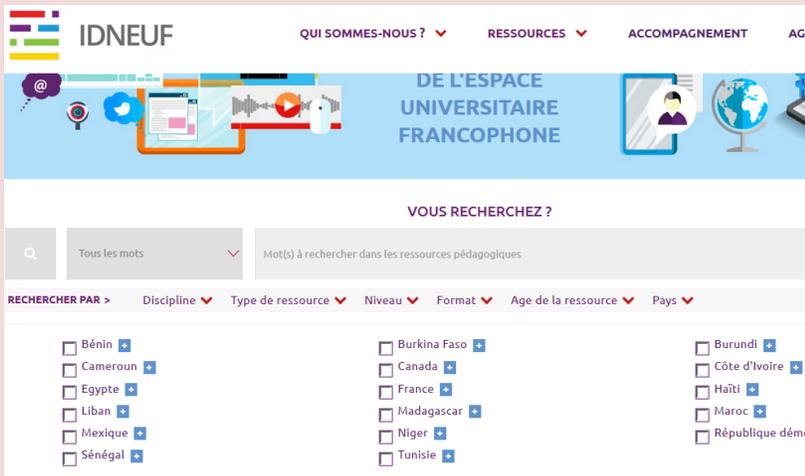 |
| *Objectifs* | - Préparer des méthodes d'accès aux ressources pédagogiques du portail francophone à partir des sites institutionnels du BAP (CNF, CNFp) ;<br>- Envisager une équipe permanente de suivi du portail qui impliquerait a minima un documentaliste ou un spécialiste e-learning pour indexer les REL et les transmettre au portail IDNEUF ;<br>- Prévoir des formations aux utilisateurs des CNF et CNFp sur la recherche documentaire en ligne pour une bonne exploitation des ressources du portail IDNEUF. |
| *Mission cadre* | Mission C « Numérique éducatif » en collaboration avec la mission A « Enseignement supérieur et transfert d'expertises » |
| *Prérequis* | - Connaitre le fonctionnement du portail IDNEUF ;<br>- Connaitre les techniques documentaires et les TIC ;<br>- Avoir une culture des licences libres et des *Creative commons* ; |
| *Responsabilité* | Équipe d'enquête des CNF et CNFp du BAP sous responsabilité d'un chargé de mission « Patrimoine pédagogique numérique » |
| *Public cible* | - Institutions universitaires publiques et privées ;<br>- Centres de formation ;<br>- Bibliothèques universitaires ;<br>- Bibliothèques de recherche ;<br>- Etc. |
| *Date de démarrage* | - Mars 2017 |
| *Durée* | - 6 mois |
| *Produits escomptés* | - Guide utilisateur de l'exploitation des REL ; |
| *Types de formations nécessaires* | - Exploiter les formations proposées sous l'activité A et B<br>- Formation sur la recherche documentaire et l'accès aux RELs. |
| *Type(s) de compétences(s) associées (Référentiel AUF)* | - (C1073) Formuler une question de recherche sur la base de sa connaissance de la littérature du domaine et de sa pratique du terrain<br>- (C1029) Concevoir et mettre en œuvre un dispositif de recherche permettant de répondre à la question posée<br>- (C1023) Communiquer le résultat de ses recherches<br>- (C1130) Travailler, échanger, se divertir, rechercher l'information avec des applications nomades<br>- (C1061) Élaborer une stratégie de recherche d'information<br>- (C1090) Localiser l'information recherchée sur Internet<br>- (C1125) Synthétiser les résultats de recherche d'informations sur Internet<br>- (C1112) Rechercher l'information |
| *Coûts prévisionnel* | - Coûts à estimer selon les grilles AUF pour un chargé de mission de 3 mois supplémentaires « Patrimoine pédagogique numérique »<br>- Coûts d'un atelier de formation de 5 jours pour l'équipe chargée de la conception des modules de recherche, d'accès et de communication des RELs.. |
| *Indicateurs de qualité* | - Qualité du service de communication des RELs rassemblées ;<br>- Quantité des RELs transmises au serveur IDNEUF ;<br>- Degré de visibilité des ressources du BAP sur le serveur IDNEUF (qualité de l'indexation) |

### 2.1.4. Ressources pédagogiques : création de cours en ligne

| | |
|---|---|
| *Type d'activité* | ***Création de cours en ligne structurés et interopérables*** |
| *Argumentaire* | La numérisation des cours et leur mutualisation en ligne n'est pas encore à son optimum dans le contexte francophone en général. Les enseignants ont tendance à produire des versions numériques de leurs supports de cours en format DOC ou PDF ou PPT qui sont des formats fermés et peu structurés pour la réutilisabilité et la mutualisation. |
| *Objectifs* | La conception de cours en ligne structurés et interopérables répond de plus en plus à des critères normatifs et des techniques de granularité et d'agrégation peu maîtrisées par les enseignants. Il est important pour la réalisation d'un patrimoine éducatif numérique de former les enseignants à la conception structurée à la source de supports pédagogiques numériques.<br>- Préparer un cadre opérationnel pour commander auprès de personnes et des institutions la numérisation de cours ;<br>- Aider les institutions universitaires à numériser leurs matériels pédagogique ;<br>- Préparer des enseignants formateurs à la conception interopérables de leurs ressources pédagogiques ;<br>- Instaurer au sein des institutions d'enseignements une culture de conception normalisée de ressources pédagogiques interopérables ;<br>- Prévoir des formations à la fois aux utilisateurs des CNF et CNFp et des universités (formations hybrides) pour l'enrichissement des techniques de conception/création de ressources pédagogiques interopérables,<br>- Enrichir la qualité des RELS à transmettre au portail IDNEUF et faciliter/accélérer leur intégration ; |
| *Mission cadre* | Mission C « Numérique éducatif » en collaboration avec la mission A « Enseignement supérieur et transfert d'expertises » |
| *Prérequis* | - Disposer d'un cahier des charges et un modèle de convention pour motiver les enseignants à numériser et déposer leurs supports de cors selon des procédures auxquelles ils auraient été formés ;<br>- Avoir déjà établi des accords-cadres avec des institutions pour numériser et restituer leurs ressources pédagogiques numérisées ; |
| *Responsabilité* | Équipe d'enquête des CNF et CNFp du BAP sous responsabilité d'un chargé de mission « Patrimoine pédagogique numérique » |
| *Public cible* | - Enseignants universitaires ;<br>- Formateurs ;<br>- Producteurs de documents scientifiques (thèses, mémoires, etc.)<br>- Etc. |
| *Date de démarrage* | - Novembre 2017 |
| *Durée* | - 12 mois |
| *Produits escomptés* | - Stratégie de production sous licence *Creative Commons* de cours en ligne structurés et interopérables ;<br>- Cahier des charges pour négocier la création de cours en ligne structurés et interopérables ;<br>- Guide utilisateur de conception / structuration de cours en ligne structuré et interopérables ; |
| *Types de formations nécessaires* | - Formation sur la création de cours en ligne structurés et interopérables ; |
| *Type(s) de compétences(s) associées (Référentiel AUF)* | - (C1058) Éditer des contenus sur l'Internet<br>- (C1009) Indexation des Ressources Pédagogiques Libres<br>- (C1127) Traiter, gérer et organiser l'information numérique<br>- (C1115) Rendre des ressources Web disponibles sous une licence libre |

| *Coûts prévisionnel* | - Coûts à estimer selon un modèle économique intégrant le coût du développement d'un module de cours en ligne |
|---|---|
| *Indicateurs de qualité* | - Nombre de cours produits ;<br>- Nombre de répondants à l'appel à projets de cours en ligne ;<br>- Qualité des rapports de l'activité ;<br>- Temps de réalisation des objectifs. |

Les activités autour des RELs et des cours en lignes sont multiples, en amont et en aval du processus de leur acquisition, traitement et mise à disposition. Elles engagent des choix cohérents de formation (sur des compétences réelles et productives), de développement (de systèmes et applications numériques innovants) et de procédures d'exploitation et de communication (un système de veille stratégique).

L'étude à propos du contexte éducatif de la région de l'Asie pacifique démontre qu'il y a un besoin réel de recadrer la formation francophone sur les TIC/E pour l'orienter sur les voies de l'innovation et de la cohérence. L'un des objectifs de la mission A du BAP est de « mobiliser une offre d'expertise francophone à travers les pôles scientifiques régionaux dans le cadre d'une approche par projet ». Cette démarche est fort importante pour qu'elle soit généralisée à toutes les missions en vue d'optimiser les moyens et les ressources disponibles. Toutes les formations devraient impérativement être liées à une programmation claire avec des objectifs précis et non pas à des choix sectoriels, aléatoires et isolés. Chaque formation devrait généralement figurer dans le cahier des charges d'un programme ou d'un projet et devrait justifier de son utilité dans la conception globale de ce programme ou projet.

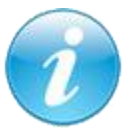

> Le simple fait de répondre à un besoin de formation exprimé ou proposé sans une vision d'ensemble ou une articulation avec d'autres types d'actions, serait un gâchis de temps et d'énergie, car la popularité d'un choix ne peut le prémunir de devenir figé, statique et non productif..

Dans le point suivant, il est question de proposer dans la feuille de route 2016-2017, un ensemble de formations liées aux activités définies, elles-mêmes partagées entre les deux missions A et C, la première à destinations des acteurs universitaires francophones dans leurs programmes de formation LMD et la deuxième à destination des acteurs individuels ou institutionnels francophones publics ou privés (enseignants, formateurs, producteurs de contenus pédagogiques, personnel administratif, ingénieurs d'études, etc.).

## 2.2. Des activités de formation et d'accompagnement relatives au renforcement des capacités autour du numérique éducatif francophone en Asie pacifique

La question qui se pose de prime abord est : comment et sur quelle base définir des compétences types pour proposer des accompagnements (des ateliers ou des parcours de formation thématiques) capables de renforcer les capacités TIC/E dans la région du BAP ?

Deux démarches sont mises en application :

1) d'abord une lecture dans l'état des lieux des formations francophones recensées par l'AUF et l'IFIC qui a permis d'identifier des déséquilibres importants dans la répartition des thématiques des TIC/E entre les régions francophones de l'AUF. L'idée principale de cette lecture n'est pas de proposer des solutions de formation qui résorbent entièrement le déséquilibre observé. L'idée est surtout de prévoir des parcours de formation pour des compétences types qui pourraient avoir de fortes chances d'être sollicitées par les partenaires francophones de la région.

2) ensuite, une lecture dans les résultats de l'enquête effectuée dans les trois pays CLV (voir partie 1) qui a permis de mette en évidence une réalité à double facettes. Une première facette dévoile l'état précaire de l'existant en termes de ressources et de services TIC/E dans la région de l'Asie pacifique. Une deuxième facette reflète les attentes souvent ambitieuses exprimées par les différents acteurs rencontrés lors des visites de terrain.

Le croisement entre les deux démarches a permis de définir des domaines de formation clés qui ont été des parents pauvres dans l'historique des formations francophones en Asie pacifique. Pourtant, ces compétences « professionnalisantes » constituent des catalyseurs d'innovation qu'il faudrait impérativement programmer dans le contexte pédagogique régional :

- les REL (conception, référencement et accès) ;
- les licences libres et le libre accès ;
- la structuration des cours en ligne et la scénarisation pédagogique;
- la démarche qualité ;
- la gestion de projets d'offres de formation en ligne ;
- l'apprentissage mobile ;
- les CLOM.

Ces compétences pourraient bien entrer dans le cadre d'une activité d'accompagnement d'élèves et enseignants pour les former et les certifier à l'usage des TIC/E.

| Type d'activité | *Accompagnement d'élèves et enseignants pour concevoir et réaliser des dispositifs et outils de formation* |
|---|---|
| *Argumentaire* | - |
| *Objectifs* | - *Formation puis certification des enseignants, instituteurs, professeurs des collèges et lycées à l'utilisation des technologies dans l'éducation :*<br>- *Maîtrise, conception et réalisation d'outils, de contenus et de dispositifs d'apprentissage* : engage une série d'actions du type : Cela engage des actions d'Analyse et structuration des contenus de cours selon l'unité d'organisation type de l'établissement (cours, UV, crédits, module, etc. 4) (scénario de l'ensemble du cours) ; Définition des objectifs (d'intégration, généraux objectifs opérationnels) pour le cours mais aussi pour chaque unité d'enseignement/apprentissage et pour chaque activité au sein de celle-ci ; Scénarisation aux trois niveaux de structuration en cohérence avec les différents niveaux des objectifs (unité d'enseignement et activité au sein de celle-ci) et les contenus ; |
|  | Choix des approches pédagogiques pour chaque activité d'enseignement/apprentissage ; Identification des ressources et les activités à médiatiser ; Choix, à chacun des niveaux du cours (une activité, une séance, le cours dans son entièreté), des dispositifs TIC/TICE (internes à la plateforme, éventuellement externes) les plus adaptés en fonction des objectifs, de la discipline, du public, des contextes organisationnel et technique, etc. ; Réalisation par soi ou par un tiers des ressources nécessaires (documents, activités d'apprentissage numériques médiatisées textuelles et multimédias, capsules vidéo, animations, simulations, etc.) ; Conception de scénario d'accompagnement et de soutien en plus des scénarios d'apprentissage ; Prévision des modalités d'évaluation des apprentissages (atteinte des objectifs).<br>- *Développer une culture du Design et des Usages* : Dans les universités ces compétences en design deviennent essentielles pour les acteurs opérant au sein des cellules TICE des universités : design pédagogique, design des contenus multimédias, design des interactions avec les usagers, design des e-services. Les apports de connaissances sur le design permettent en effet des processus de conception qui soient plus efficaces, plus efficients, plus ergonomiques, plus acceptables socialement, plus attractif voire ludique. Il est un facteur de réussite11, de réduction des résistances et d'adoption par les usagers et d'inclusion pour des publics ayant des caractéristiques différences. C'est le design inclusif. |
| *Mission cadre* | Mission C « Numérique éducatif » en collaboration avec la mission A « Enseignement supérieur et transfert d'expertises » |
| *Prérequis* | - Maîtriser un minimum d'applications informatiques du domaine bureautique, multimédia et d'Internet ;<br>- Avoir de l'expérience dans l'activité d'enseignement et de formation ;<br>- Être proposé à une fonction liée à l'enseignement et la formation |
| *Responsabilité* | Mission C « Numérique éducatif » en collaboration avec la mission A « Enseignement supérieur et transfert d'expertises |
| *Public cible* | - Enseignants des institutions scolaires et universitaires publiques et privées ;<br>- Spécifiquement des élèves-enseignants de l'institut national de l'éducation, inscrits au CLOM CERTICE-SCOL de l'AUF (proposé par l'antenne de Phnom Penh en partenariat avec l'IFIC et l'Université Cergy Pontoise),<br>- Etc. |
| *Date de démarrage* | - Octobre 2017 |
| *Durée* | - 3 mois |
| *Produits escomptés* | - Document de stratégie d'accompagnement à la formation TIC/E<br>- Programmes/parcours types de formations sur les TIC/E ;<br>- Supports de formation appropriés ; |
| *Types de formations nécessaires* | - Exploiter les formations proposées sous les autres activités ;<br>- Formation sur les différentes compétences liées à l'usage des outils TIC/E, contenus pédagogiques et modèles pédagogiques d'apprentissage. |
| *Type(s) de compétences(s) associées (Référentiel AUF)* | - (C1007) Analyser les besoins et les traduire en objectifs d'apprentissage<br>- (C1119) Scénariser un parcours de formation de type CLOM/MOOC<br>- (C1120) Structurer le dispositif de formation et concevoir le scénario d'apprentissage<br>- (C1032) Concevoir le scénario d'accompagnement et de soutien des apprenants<br>- (C1033) Concevoir les modalités d'évaluation des acquis des apprenants<br>- (C1063) Élaborer une stratégie d'évaluation du dispositif de formation<br>- (C1066) Évaluer le dispositif de formation en vue d'en améliorer la qualité<br>- |
| *Coûts prévisionnel* | - Modèle économique selon les grilles AUF incluant des frais d'inscription mais aussi des prises en charge pour la motivation ;<br>- Coûts de gestion d'un atelier de formation de 5 jours ; |
| *Indicateurs de qualité* | - Qualité des rapports et documents de travail produits ;<br>- Taux de sensibilisation auprès des structures visées ;<br>- Qualité et quantité des personnes ayant été touchés par l'activité. |

Dans l'étape suivante, ces besoins en formations TIC/E seront traduits en compétences à partir du nouveau référentiel TIC/E de l'AUF et programmées dans l'agenda de formation du BAP pour 2016-2017. Les compétences seront extraites avec leurs modules dépendants. Ces modules sont du type générique comme les défis numériques (D), les métiers du numérique (M) ou les activités pédagogique (A). Chaque compétence est décomposée en trois modules spécifiques sous forme de Savoirs (S), savoir-faire (SF) et savoir-être (SE). Néanmoins, par économie du détail ces trois modules spécifiques ne seront pas détaillés ici. Il est toutefois possible de les consulter en ligne sur le site du référentiel AUF au même titre que les définitions des autres modules du référentiel.

Mais rappelons au préalable certains principes dans le mode opératoire des formations de l'AUF :

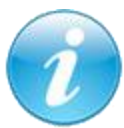

Une formation AUF n'est pas stéréotypée. Si le modèle dominant est celui du programme Transfer, une formation peut aussi varier en durée (d'un jour à des mois, voire permanente). Elle peut aussi avoir des formes pédagogiques différentes (parcours de formation, atelier, formation diplômante), et suivre des modèles pédagogiques différents (Barcamp, Classe inversée, Hybride, entièrement à distance, CLOM, …).

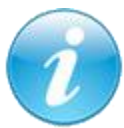

Une formation est tributaire de plusieurs facteurs : du type de public cible, des ressources pédagogiques utilisées, de la durée envisagée, des modes de suivi et d'évaluation programmés, des objectifs à atteindre, etc.

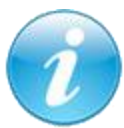

Une offre de formation aussi utile soit-elle, ne doit pas s'enfermer dans l'inertie des demandes répétitives des usagers. Elle doit aussi constituer une force de proposition pour propulser ces mêmes usagers vers des paliers de connaissances avancées, d'un usage évolué et de besoins pédagogiques renouvelés.

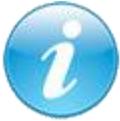

Un forfait d'atelier de 5 jours avec des formateurs locaux et des apprenants locaux couterait a minima 1.000 €. Le même atelier couterait a maxima 6.000 € quand il est réalisé avec des formateurs et des apprenants étrangers (cas très rares). Il faudrait comptabiliser donc sur une moyenne de 3.000 € par atelier, ce qui constitue une médiane pour un atelier de formation animé par des formateurs étrangers et des apprenants locaux (version courante).

## 2.2.1. Accompagnement de la communauté universitaire francophone à travers des services de formation développés dans les CNF et CNFp

À destination de candidats individuels (enseignants, étudiants, personnel administratif, ingénieurs d'études, etc.), ces formations (souvent réalisés dans les locaux des CNF et CNFp), peuvent prendre des forme variées (ateliers ou parcours de formation, Barcamp, Classe inversée, FOAD Hybride ou entièrement à distance, CLOM, …) et adressées à des profils de publics différents (des compétences professionnalisantes pour des praticiens de la FOAD ou des compétences de gouvernance pour la sensibilisation des responsables décideurs).

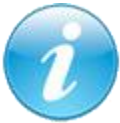

La programmation de ces ateliers devrait prévoir dans une phase initiale des regroupements régionaux d'ingénieurs de formation ou de profils assimilés (enseignants, techniciens, ingénieurs d'étude, formateurs des CNF, CNFp, etc.) qui assureront la reproduction ultérieure de ces formation à l'échelle des pays de la région.

Les formations suivantes pour les praticiens de la FOAD (professionnalisantes) sont proposées à titre indicatif. Le BAP saura ensuite en adopter, adapter, transformer les critères de toutes ou de certaines selon ses priorités et ses moyens.

**A. PARCOURS DE FORMATION : « CRÉATION DE COURS EN LIGNE, STRUCTURES & INTEROPÉRABLES »**

| *Titre(s) de compétence(s)* | - *(C1058)* **Éditer des contenus sur l'Internet** |
|---|---|
| *Définition* | L'objet de ce cours est d'aborder les points clés qu'un développeur de contenus est censé connaître (et pratiquer) pour assurer l'intégration des ressources pédagogiques dans un environnement de mutualisation via les solutions FAD. Ceci n'implique pas forcément la dimension pédagogique qui demeure plutôt du ressort des tuteurs et des enseignants. Les questions abordées ici sont plutôt d'ordre technique pour la conception matérielle et l'organisation logique des ressources en vue de faciliter leur réutilisation et leur exploitation dans différentes parcours d'apprentissage. Les cours en ligne sont de plus en plus structurés de façon modulaire fondée sur le principe de la granularité et des agrégations adaptés à des niveaux, modules et parcours de formation variés. Des outils logiciels libres sont déjà à la portée de tous proposant des méthodes simples et conviviales de découpages de contenus que les acteurs de la FAD sont en mesure de s'approprier et d'appliquer. La structuration des ressources pédagogiques se traduit en un consensus international de plus en plus large autour de la nécessité |

|   | d'harmoniser (et normaliser) l'agrégation des contenus d'apprentissage en ligne selon des principes de la modularité.<br><br>Parmi les savoirs nécessaires pour cette compétence :<br>- Connaître les bonnes pratiques d'écriture sur Internet ;<br>- Connaître les outils de gestion de contenu ;<br>- Connaître les moyens de référencement ;<br>- Connaître les bases des technologies Web ;<br>- Connaître les outils de gestion de contenu. |
|---|---|
| *Mission associée* | - Mission C « Numérique éducatif » en collaboration avec la mission A « Enseignement supérieur et transfert d'expertises |
| *Public cible* | - Enseignants ; formateurs, producteurs de ressources pédagogiques ; |
| *Lieu(x) (accueil)* | - Ho-Chi-Minh (regroupement régional d'ingénieurs de formation) |
| *Date(s)* | - Janvier 2017 |
| *Durée* | - 5 jours |
| *Type de formation* | - Présentiel |
| *Coût prévisionnel* | - Selon grille locale |
| *Indicateurs* | - Qualité des fils rouges (projets individuels) |

**B. PARCOURS DE FORMATION : « INDEXATION DES RELS »**

| *Titre(s) de compétence(s)* | - **(C1009) : *Indexation* des Ressources Pédagogiques Libres** |
|---|---|
| *Définition* | L'objet de ce parcours est d'aborder les points clés qu'un développeur de contenus est censé connaître (et pratiquer) pour assurer l'intégration des ressources pédagogiques dans un environnement de mutualisation via les solutions FAD. Ceci n'implique pas forcément la dimension pédagogique qui demeure plutôt du ressort des tuteurs et des enseignants. Les questions abordées ici sont plutôt d'ordre technique pour la description matérielle et logique des ressources, de leur référencement et de leur exploitation dans des environnements ouverts et distribués. Ces ressources sont soumises de plus en plus à des processus d'indexation par les auteurs et les professionnels de l'information. Des solutions en logiciels libres sont déjà à la portée de tous proposant des méthodes simples e conviviales de description que les acteurs de la FAD sont en mesure de s'approprier et d'appliquer. L'indexation des ressources pédagogiques se traduit en un consensus international de plus en plus large autour de la nécessité d'harmoniser (et normaliser) la description et le référencement des contenus d'apprentissage en ligne. Des référentiels normatifs de métadonnées pédagogiques sont de plus en plus utilisés par les concepteurs de cours en ligne et les développeurs de plates-formes pédagogiques. |
| *Mission associée* | - Mission C « Numérique éducatif » en collaboration avec la mission A « Enseignement supérieur et transfert d'expertises |
| *Public cible* | - Ingénieurs de formation et/ou profil assimilé (enseignants, formateur CNF, etc.) |

| | | |
|---|---|---|
| *Lieu(x) (accueil)* | - | Hanoï (regroupement régional d'ingénieurs de formation) |
| *Date(s)* | - | Décembre 2016 |
| *Durée* | - | 5 jours |
| *Type de formation* | - | Présentiel |
| *Coût prévisionnel* | - | Selon grille locale |
| *Indicateurs* | - | Qualité des fils rouges (projets individuels) |

**C. PARCOURS DE FORMATION : « LICENCE LIBRE »**

| | | |
|---|---|---|
| *Titre(s) de compétence(s)* | - | *(C1056)* **Différencier les licences ouvertes et fermées** |
| | - | *(C1115)* **Rendre des ressources Web disponibles sous une licence libre** |
| *Définition* | - | Distinguer entre les licences ouvertes et fermées des ressources et leurs droits de réutilisation dans un contexte pédagogique. |
| | - | Remixer des ressources disponibles sur Internet en licence libre. |
| *Mission associée* | - | Mission C « Numérique éducatif » en collaboration avec la mission A « Enseignement supérieur et transfert d'expertises » |
| *Public cible* | - | Ingénieurs de formation et/ou profil assimilé (enseignants, formateur CNF, etc.) |
| *Lieu(x) (accueil)* | - | Phnom-Penh (regroupement régional d'ingénieurs de formation) |
| *Date(s)* | - | Février 2017 |
| *Durée* | - | 5 jours |
| *Type de formation* | - | Présentiel |
| *Coût prévisionnel* | - | Selon grille locale |
| *Indicateurs* | - | Qualité des fils rouges (projets individuels) |

**D. PARCOURS DE FORMATION : « FONDAMENTAUX DE LA SCÉNARISATION PÉDAGOGIQUE»**

| | | |
|---|---|---|
| *Titre(s) de compétence(s)* | - | *(C1118)* **Scénariser un dispositif d'e-formation à chacun de ses niveaux de granularité (dispositif de formation, séance de cours, activité, etc.)** |
| | - | *(C1117)* **Scénariser des séquences vidéo pédagogiques** |
| *Définition* | - | La formation vise à maîtriser les concepts fondamentaux, les modèles et les outils de la scénarisation pédagogique en vue de : <br>▪ concevoir des formations efficaces en articulant de façon structurée et pertinente les connaissances visées et les activités proposées aux apprenants ; <br>▪ développer sa capacité à innover en intégrant de nouvelles approches pédagogiques (la classe inversée, la démarche par projets, l'apprentissage par le jeu, etc.), de nouvelles pratiques sociales (les cours massifs en ligne, les |
| | | réseaux sociaux, la mobilité), ou de nouvelles technologies matérielles (ex. : la réalité augmentée) ;<br>▪ anticiper la mise en œuvre opérationnelle des scénarios conçus sur des plateformes de formation en ligne |
| *Mission associée* | - | Mission C « Numérique éducatif » en collaboration avec la mission A « Enseignement supérieur et transfert d'expertises » |
| *Public cible* | - | Ingénieurs de formation et/ou profil assimilé (enseignants, formateur CNF, etc.) |
| *Lieu(x) (accueil)* | - | Vientiane (regroupement régional d'ingénieurs de formation) |
| *Date(s)* | - | Avril 2017 |
| *Durée* | - | 5 jours |
| *Type de formation* | - | Présentiel |
| *Coût prévisionnel* | - | Selon grille locale |
| *Indicateurs* | - | Qualité des fils rouges (projets individuels) |

**E. PARCOURS DE FORMATION : « ADMINISTRATION D'UNE PLATEFORME FOAD »**

| | | |
|---|---|---|
| *Titre(s) de compétence(s)* | - | *(C1087)* **Installer et administrer une plate-forme de formation massive et en ligne** |
| | - | *(C1089)* **Intégrer un contenu pédagogique au sein d'une plate-forme de formation en ligne et massive** |
| *Définition* | - | Maîtriser le processus d'installation d'une plate-forme de formation massive, de son administration et de sa maintenance ; |
| | - | Maîtriser les bases de l'utilisation d'une plate-forme pour mettre en ligne du contenu pédagogique ainsi que la gestion des principales fonctionnalités. S'appuyer pour cela sur son expertise d'enseignant en présence et à distance, sur les concepts présentés pendant la formation et sur les tâches qu'il y aura réalisées sur son propre projet de MOOC/CLOM. |
| *Mission associée* | - | Mission C « Numérique éducatif » en collaboration avec la mission A « Enseignement supérieur et transfert d'expertises » |
| *Public cible* | - | Ingénieurs de formation et/ou profil assimilé (enseignants, formateur CNF, etc.) |
| *Lieu(x) (accueil)* | - | Danang (regroupement régional d'ingénieurs de formation) |
| *Date(s)* | - | Juin 2017 |
| *Durée* | - | 5 jours |
| *Type de formation* | - | Présentiel |
| *Coût prévisionnel* | - | Selon grille locale |
| *Indicateurs* | - | Qualité des fils rouges (projets individuels) |

## F. PARCOURS DE FORMATION : « ANIMER ET PILOTER UN MOOC »

| | | |
|---|---|---|
| *Titre(s) de compétence(s)* | - | **(1010) Animer et piloter un parcours de formation de type CLOM/MOOC** |
| | - | **(C1119) Scénariser un parcours de formation de type CLOM/MOOCs** |
| *Définition* | - | Préparer l'intégration d'un CLOM dans un cursus de formation initiale (exemple de l'Institut national de l'Éducation) ; |
| | - | Maîtriser les bases de la scénarisation pédagogique d'un cours en ligne ouvert et massif (CLOM/MOOC). L'apprenant s'appuiera pour cela sur son expertise d'enseignant en présence et à distance, sur les concepts présentés pendant la formation et sur les tâches qu'il y aura réalisées sur son propre projet de MOOC |
| | - | Maîtriser les différents outils d'animation et les modes d'interactions synchrones et asynchrones, textuels et vidéos, de manière à motiver les participants. Organiser la communication et l'animation du MOOC dans la durée. S'appuyer pour cela sur son expertise d'enseignant en présence et à distance, sur les concepts présentés pendant la formation et sur son expérience de participation à des MOOC |
| *Mission associée* | - | Mission C « Numérique éducatif » en collaboration avec la mission A « Enseignement supérieur et transfert d'expertises » |
| *Public cible* | - | Ingénieurs de formation et/ou profil assimilé (enseignants, formateur CNF, etc.) |
| *Lieu(x) (accueil)* | - | Ho-Chi-Minh ville (regroupement régional d'ingénieurs de formation) |
| *Date(s)* | - | Septembre 2017 |
| *Durée* | - | 5 jours |
| *Type de formation* | - | Présentiel |
| *Coût prévisionnel* | - | Selon grille locale |
| *Indicateurs* | - | Qualité des fils rouges (projets individuels) |

Le BAP pourrait aussi prévoir des formations pour améliorer la gouvernance des offres de formation FOAD (responsables décideurs, administrateurs, etc.).

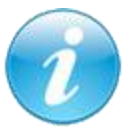

> Pour les responsables décideurs, il s'agit généralement de formation de niveau débutant, puisque l'objectif principal est de les sensibiliser, les initier et les motiver à un concept nouveau. Un parcours de formation pour les responsables décideur doit rester initiatique, en face à face, court et polyvalent.

Les formations suivantes de ce type sont proposées à titre indicatif. Le BAP saura ensuite en adopter, adapter, transformer les critères de toutes ou de certaines selon ses priorités et ses moyens.

## G. PARCOURS DE FORMATION : « GOUVERNANCE D'UN SYSTÈME D'INFORMATION »

| | | |
|---|---|---|
| *Titre(s) de compétence(s)* | - | **(C1013) Assurer la gouvernance d'un SI** |
| *Définition* | - | Définir, déployer et contrôler la gestion des systèmes d'information en ligne avec les ambitions de la partie prenante. Prendre en compte tous les paramètres internes et externes tels que la conformité aux normes légales et industrielles afin d'orienter la gestion du risque et le déploiement de ressources pour améliorer le niveau de service à la partie prenante. Parmi les objectifs :<br>▪ Faire des participants des acteurs importants dans l'ouverture de la culture scientifique au plus grand nombre.<br>▪ Améliorer sa gouvernance au travers de l'informatique décisionnelle, du Big-data et de l'intelligence collective. |
| *Mission associée* | - | Mission C « Numérique éducatif » en collaboration avec la mission A « Enseignement supérieur et transfert d'expertises » |
| *Public cible* | - | Acteurs éducatifs responsables (Doyens, chefs de départements, chefs d'entreprises, responsables CNF, etc.) |
| *Lieu(x) (accueil)* | - | Ho-Chi-Minh Ville (regroupement régional de responsables de formation) |
| *Date(s)* | - | Décembre 2016 |
| *Durée* | - | 2 jours |
| *Type de formation* | - | Présentiel |
| *Coût prévisionnel* | - | Selon grille locale |
| *Indicateur* | - | Taux de satisfaction des participants |

## H. PARCOURS DE FORMATION : « MÉDIATION DANS UN ESPACE D'INNOVATION »

| | | |
|---|---|---|
| *Titre(s) de compétence(s)* | - | **(C1016) Assurer la médiation numérique spécialisée dans la fabrication numérique**<br>- **(C1046) Définir les objectifs généraux et spécifiques d'un dispositif d'e-e-formation**<br>- **(C1050) Développer de nouvelles valeurs d'usage pour la création d'une médiation dans un espace d'innovation**<br>- **(1096) Mettre en œuvre le partenariat indispensable à la mise en œuvre d'un dispositif d'e-formation** |
| *Définition* | - | Mettre en place des techniques d'identification et de création de valeurs communes autour de projets numériques ;<br>- Connaître les concepts liés à l'identification de valeurs communes pour le management de l'innovation numérique (éthique, relation entre impulsion et disposition, conception, expérimentation, durabilité)<br>- Connaître les processus collectifs de développement d'une idée, de modalités de projet d'apprentissage ou de recherche par l'innovation |
| | - | Mobiliser des ressources humaines complémentaires pour la création d'un éco système innovant par le numérique |
| | - | Connaître les outils, les techniques de management de l'innovation et de la créativité (anticipation, imaginaire, nouveaux territoires) |
| *Mission associée* | - | Mission C « Numérique éducatif » en collaboration avec la mission A « Enseignement supérieur et transfert d'expertises » |
| *Public cible* | - | Acteurs éducatifs responsables (Doyens, chefs de départements, chefs d'entreprises, responsables CNF, etc.) |
| *Lieu(x) (accueil)* | - | Phnom-Penh (regroupement régional de responsables de formation) |
| *Date(s)* | - | Janvier 2017 |
| *Durée* | - | 2 jours |
| *Type de formation* | - | Présentiel |
| *Coût prévisionnel* | - | Selon grille locale |
| *Indicateur* | - | Taux de satisfaction des participants |

**I. PARCOURS DE FORMATION : « GESTION DE L'INNOVATION »**

| | | |
|---|---|---|
| *Titre(s) de compétence(s)* | - | *(C1108)* **Promouvoir l'innovation** |
| | - | *(C1095)* **Mettre en correspondance innovations, besoins pédagogiques, usages** |
| | - | *(C1121)* **Structurer un espace d'innovation technologique de façon participative** |
| | - | *(C1050)* **Développer de nouvelles valeurs d'usage pour la création d'une médiation dans un espace d'innovation** |
| *Définition* | - | Envisager des solutions créatives pour fournir de nouveaux concepts, idées, produits ou services. Promouvoir une pensée ouverte et innovante pour exploiter les avancées technologiques dans les besoins ou la définition des objectifs des parties prenantes |
| | - | Mobiliser des ressources humaines complémentaires pour la création d'un éco système innovant par le numérique |
| | - | Mettre en place des techniques d'identification et de création de valeurs communes autour de projets numériques |
| *Mission associée* | - | Mission C « Numérique éducatif » en collaboration avec la mission A « Enseignement supérieur et transfert d'expertises » |
| *Public cible* | - | Acteurs éducatifs responsables (Doyens, chefs de départements, chefs d'entreprises, responsables CNF, etc.) |
| *Lieu(x) (accueil)* | - | Hanoï (regroupement régional de responsables de formation) |
| *Date(s)* | - | Mars 2017 |
| *Durée* | - | 2 jours |
| *Type de formation* | - | Présentiel |
| *Coût prévisionnel* | - | Selon grille locale |
| *Indicateur* | - | Taux de satisfaction des participants |

La proposition de ces parcours de formation devrait être vérifiée avec l'IFIC pour éviter de la redondance avec des parcours déjà existants. Les parcours recensés sur le site de l'IFIC sont à cette date (mai 2016) au nombre de 21. Leur nombre va crescendo au fur et à mesure de la mise en application du nouveau référentiel de compétences TIC/E.

L'AUF a déjà commencé à recenser auprès de ses établissements membres, de nouveaux besoins en formation. Dans ce contexte et en rapport avec l'axe "Formation et sensibilisation" de l'IFIC, plusieurs parcours de formation ont été sélectionnés suite à un appel à candidatures pour la conception de ressources pédagogiques dans le domaine des TIC/E. Si le BAp devrait définir les contenus de nouveaux parcours de formation, ces contenus seraient déposés auprès e lIFIC et répertoriés pour servir à d'autres formations similaires dans d'autres contextes de formations francophones.

La feuille de route prévoit toutefois de proposer des parcours de formation qui seraient utiles pour la mission A « Enseignement supérieur et transfert d'expertise ».

### 2.2.2. *Appui au développement de projets innovants dans le cadre de la mission A « Enseignement supérieur et transfert d'expertise » à destination des acteurs de la francophonie universitaire de la région Asie pacifique.*

Parmi les objectifs de la mission A, il est prévu « *accompagner les différents établissements membres de la région qui en font la demande, dans leur volonté de mettre en place des projets interuniversitaires de formation francophone de niveau licence et de niveau master* ». En d'autres termes, le BAP accompagnera les universités qui le demanderaient à déployer leurs systèmes FOAD aux niveaux de la Licence et du Master et renforcer les capacités de la recherche au niveau du Doctorat. Ceci impliquerait, entre autres, plusieurs types et niveaux de formations à la fois du personnel enseignant et des étudiants de ces universités pour qu'ils puissent installer, utiliser, gérer, administrer et évaluer le dispositif FOAD de leur université.

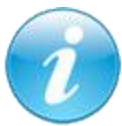

L'administration d'un dispositif FOAD étant parfois complexe dans ses différentes composantes technologique et pédagogique, la feuille de route proposerait des parcours de formation standards que les responsables de la mission A, en collaboration avec ceux de la mission C, pourraient adapter et proposer à leurs universités partenaires.

Parmi les compétences qui iraient mieux à la fois avec les exigences d'un mode d'enseignement hybride (plus approprié pour une formation universitaire) et les orientations stratégiques de la mission A, on pourrait proposer des compétences comme :

- Déploiement d'une plateforme de « eformation » : installation, administration et intégration de contenu pédagogique - exemple et cas pratique : Moodle
- Tutorat à distance
- Évaluation dans les dispositifs d'apprentissage

- Organisation d'une classe/un cours inversé avec des outils pertinents et des applications en ligne adéquates
- Passage d'un enseignement conventionnel à l'apprentissage multimodal mobile (la classe inversée enrichie)
- Intégration des ressources sur une PF de MOOC et gestion des fonctionnalités (Open EdX)
- Ré-enchantement de la transmission des savoirs, savoir-faire, savoir-être au moyen des jeux vidéo à vocation pédagogique dans un contexte d'enseignement numérique

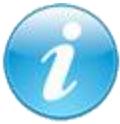

L'avantage de ces thèmes de formation est qu'ils font déjà l'objet de parcours programmés par l'IFIC et donc leurs ressources pédagogiques sont déjà existantes et validées. L'IFIC pourrait fournir les modèles de montages et les contenus appropriés à ces parcours de formation.

Le BAP pourrait envisager la répartition suivante :

| Lieu | Thème | Date |
|---|---|---|
| Hanoï | - Gérer la qualité informatique d'un SI<br>- Intégration des ressources sur une PF de MOOC et gestion des fonctionnalités (Open EdX) | Mai 2017<br>Novembre 2017 |
| Danang | - Déploiement d'une plateforme de « eformation » : installation, administration et intégration de contenu pédagogique - exemple et cas pratique : Moodle | Janvier 2017 |
| Ho-Chi-Minh | - Tutorat à distance | Mars 2017 |
| Phnom-Penh | - Classe inversée | Juin 2017 |
| Vientiane | - Évaluation dans les dispositifs d'apprentissage | Octobre 2017 |

### 2.2.3. *Appui au développement d'offres de FORMATIONS HYBRIDES dans le cadre de la mission A « Enseignement supérieur et transfert d'expertise » à destination des acteurs de la francophonie universitaire de la région Asie pacifique.*

La formation hybride se développe de plus en plus dans le discours institutionnel mais aussi dans la pratique, pour différentes raisons (coût, souplesse, mais aussi intérêt didactique).

L'accompagnement des universités francophones par la FOAD dans leurs programmes de formation académique (Licence, Master) passera inéluctablement par des solutions du type Hybride : association à des pourcentages variés entre le présentiel et la distance. Ces proportions sont à étudier selon le contexte de la formation et les moyens mises à disposition qui aboutiront à des formes différentes d'hybridation :

- présentiel enrichi par des supports multimédias ;
- présentiel amélioré par du travail en amont ou en intersession ;
- présentiel alterné ;
- présentiel allégé ou réduit.

La mise en place de chacune de ces solutions hybrides n'est pas sans soulever de questions ni poser des problèmes. Cette modalité renvoie à des questions clés : personnalisation des parcours de formations, posture de formateur, place de l'outil technique dans la formation, tutorat et modes d'évaluation, etc.

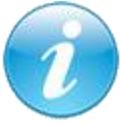

> Avant de se lancer dans le choix d'un type de formation hybride, il est important de considérer les connaissances et les compétences à développer ainsi que la situation des apprenants pour sélectionner la meilleure option selon le contexte. Chacun des types de formation a ses forces et des stratégies qui lui sont propres tout comme ses limites. Il est donc essentiel de ne pas faire ce choix à la légère ou selon la tendance du moment.

Aussi, pour anticiper toute sollicitation de la part des universités francophones pour les accompagner dans le développement de leurs stratégies en matière de pédagogie, le BAP devrait constituer une équipe de réflexion/développement qui aurait la charge de définir une politique de formations hybrides. Cette équipe devrait agir dans un cadre maîtrisé sur trois plans :

1) **Juridique et légal** : connaître les textes juridiques et réglementaires de chaque pays en rapport avec la formation à distance en vue d'identifier la validation et la reconnaissance des diplômes attribués par ce procédé pédagogique ;

2) **Technologique** : connaître l'environnement technologique de chaque pays en termes d'infrastructures de télécommunications et des lois s'y afférant. Cela concerne les réseaux de télécommunications et les réseaux numériques de transmission de données, les opérateurs Internet et les débits de leurs connections, etc.

3) **Institutionnel** : connaître les textes réglementaires des universités pour savoir leurs disponibilités juridiques à reconnaitre des diplômes dispensés via FOAD et formations hybrides, connaître le niveau d'équipement des universités pour installer des dispositifs FOAD et hybrides, savoir si elles ont le personnels adéquat pour gérer des dispositifs de ce type et si des formations préalables sont nécessaires, etc.

Une activité peut a priori être programmée par le BAP concernant la formation hybride pour préparer un cadre opérationnel interne avant de traiter avec les universités partenaires qui le souhaiteraient :

| | |
|---|---|
| *Type d'activité* | ***Conception et définition d'un modèle de formation hybride au profit des universités francophones de l'Asie pacifique*** |
| *Argumentaire* | Cette activité est proposée dans le cadre de la mission A. Elle vise à mobiliser une offre d'expertise dans la cadre d'une approche par projet (licence et master) pour un guide accompagnement d'une nouvelle offre de formation dans la région. Ce guide devrait être testé dans l'accompagnement des différents établissements membres de la région qui font la demande de construire des projets de création des cours en mode hybride pour les niveaux de licence et de master. |
| | Il faut rappeler à cet égard, que l'offre francophone d'expertise développée par la CONFRASIE sur la gouvernance universitaire vise à structurer puis développer une culture de la qualité au sein des institutions d'enseignement supérieur et de recherche qui en font la demande. Pour rappel, 69 |
| | établissements, soit 80 % des membres de la conférence, ont répondu au questionnaire en ligne enfin de définir leurs priorités dans les quels 5 institutions choisissent les projets interuniversitaire sur le numérique éducatif. Bien que ce taux est encore modeste mais cela montre un engagement de la part de ces établissements dans les projets innovants dont la création des cours en ligne. |
| *Objectifs* | - Faire l'état des lieux du contexte juridique, technologique et institutionnel universitaire en lien avec la FOAD dans les pays concernés ;<br>- La mise au point et la validation d'une typologie de dispositifs de formation hybrides pouvant intéresser les universités francophones partenaires (enquête) ;<br>- La définition d'une approche d'étude des effets sur l'apprentissage et le développement professionnel de l'institution (étude pilote sur échantillon) ;<br>- La définition d'un plan de recueil de données ;<br>- La définition des caractéristiques de dispositifs hybrides pour l'apprentissage ;<br>- La définition des caractéristiques de dispositifs hybrides pour le développement professionnel des enseignants ;<br>- La rédaction d'un modèle de cahier des charges qui tiendrait compte des caractéristiques des prototypes hybrides retenues : supports de cours hybride, rythmes des cours hybride dispensés (pourcentages entre présence et distance), indicateurs de réceptivité du public, besoins du public, moyens humains nécessaires, moyens financiers engagés, modes d'évaluation retenus, reconnaissance des diplômes, etc. |
| *Mission cadre* | Mission C « Numérique éducatif » en collaboration avec la mission A « Enseignement supérieur et transfert d'expertises » |
| *Prérequis* | - Avoir une vue globale sur le système éducatif des pays de la région ;<br>- Avoir une connaissance minimale du contexte juridique, technologique et institutionnel des pays ;<br>- Avoir l'expérience des dispositifs pédagogiques en ligne et du tutorat ;<br>- Avoir de l'expérience de gestion de projets ; |
| *Responsabilité* | Équipe de réflexion/développement sous la responsabilité d'un chargé de mission « Formations hybrides » |
| *Public cible* | - Enseignants ;<br>- Ingénieurs pédagogiques ;<br>- Ingénieurs de recherche ; |
| *Date de démarrage* | - Novembre 2016 |
| *Durée* | - 06 mois |
| *Produits escomptés* | - Résultats d'enquêtes : constat général sur les conditions juridiques, technologiques et institutionnelles du projet ;<br>- Typologies de dispositifs pédagogiques hybrides ;<br>- Guide et manuels d'installation et d'administration de formations hybrides ;<br>- Appel à projet de création de cours en mode hybride en licence ou en master pour la collaboration entre les universités francophones de la région Asie pacifique.. |

| Types de formations nécessaires | - Former une (des) personne(s) ressource(s) dans l'entourage des CNF, CNFp, notamment en rapport avec la mission A pour gérer un projet d'installation d'un dispositif hybride ; |
| --- | --- |
| | - Former une (des) personne(s) ressource(s) pour le suivi pédagogique des offres de formation hybride. |
| Type(s) de compétences(s) associées (Référentiel AUF) | - (C1006) Analyser le dispositif de formation existant, identifier les besoins émergents et les solutions pertinentes |
| | - (C1046) Définir les objectifs généraux et spécifiques d'un dispositif d'e-e-formation |
| | - (C1096) Mettre en œuvre le partenariat indispensable à la mise en œuvre d'un dispositif d'e-formation |
| | - (C1040) Construire le modèle économique et le budget d'un dispositif d'e-formation |
| | - (C1045) Définir les objectifs d'apprentissage généraux et opérationnels |
| | - (C1067) Évaluer les environnements techno-pédagogiques et ses outils disponibles et d'identifier les plus adaptés à un dispositif de e-formation et aux activités d'apprentissage |
| Coûts prévisionnel | - Coûts de formation de deux ateliers du type Transfer (3.000 x 2) |
| | - Frais d'enquête et de publication de résultats |
| Indicateur de qualité | - Exhaustivité du recensement ; |
| | - Qualité des données d'identification ; |
| | - Résultats de l'appel à projet de création de cours hybrides ; |
| | - Qualité des rapports produits. |

Cette activité implique donc la composition d'une équipe de réflexion chargée de mener une étude exploratoire sur les modèles hybrides et leur applicabilité dans la région. Il est aussi indispensable de penser à la formation de cette équipe sur les techniques et modalités pédagogiques d'une formation hybride pour qu'elle puisse ultérieurement être chargée de gérer les offres de formations hybrides avec les universités partenaires dans le cadre de la mission A.

Dans ce sens, deux types de formations peuvent être programmées pour cette équipe à partir du référentiel TIC/E de l'AUF :

- Étude et définition de son projet de dispositif d'e-formation
- Suivi pédagogique du dispositif de formation

### J. PARCOURS DE FORMATION : « ÉTUDE ET DÉFINITION DE SON PROJET DE DISPOSITIF DE E-FORMATION »

| Titre(s) de compétence(s) | - *(C1006)* **Analyser le dispositif de formation existant, identifier les besoins émergents et les solutions pertinentes** |
| --- | --- |
| | - *(C1046)* **Définir les objectifs généraux et spécifiques d'un dispositif d'e-e-formation** |
| | - *(C1096)* **Mettre en œuvre le partenariat indispensable à la mise en œuvre d'un dispositif d'e-formation** |
| | - *(C1040)* **Construire le modèle économique et le budget d'un dispositif d'e-formation** |
| | - *(C1045)* **Définir les objectifs d'apprentissage généraux et opérationnels** |

| | - *(C1067) Évaluer les environnements techno-pédagogiques et ses outils disponibles et d'identifier les plus adaptés à un dispositif de e-formation et aux activités d'apprentissage* |
|---|---|
| Définition | Cette formation envisage développer les compétences capables de permettre aux candidats d'assurer des fonctions de conception et de réalisation. Ils pourront à la suite de cette formation intervenir sur des projets de réalisation de produits, d'offres de formation ou de dispositifs nouveaux (depuis leur définition jusqu'à leur implémentation) aussi bien que sur la transformation, le déploiement ou le développement de dispositifs existants.<br><br>Ils pourront notamment procéder à la :<br><br>- Définition de projet (objectifs, approches pédagogiques, ressources, environnements, etc.).<br>- Analyse de la faisabilité du projet.<br>- Définition des objectifs pédagogiques généraux (en termes de discipline).<br>- Rédaction du cahier des charges.<br>- Si nécessaire, faire appel à la sous-traitance : définition des prestations à sous-traiter avec les services facultaires ou universitaires ;<br>- Définition de modèle économique / budget / coûts ;<br>- Établissement de plan de travail (évaluation du temps), calendrier des opérations. |
| Mission associée | - Mission C « Numérique éducatif » en collaboration avec la mission A « Enseignement supérieur et transfert d'expertises » |
| Public cible | - Acteurs éducatifs responsables (Doyens, chefs de départements, chefs d'entreprises, responsables CNF, etc.) |
| Lieu(x) (accueil) | - Hanoï (regroupement régional de membres de l'équipe) |
| Date(s) | - Février 2017 |
| Durée | - 5 jours |
| Type de formation | - Présentiel |
| Coût prévisionnel | - Selon grille locale |
| Indicateur | - Pertinence des rapports |

## K. PARCOURS DE FORMATION : « SUIVI PÉDAGOGIQUE DU DISPOSITIF DE FORMATION »

| Titre(s) de compétence(s) | - *(C1131) Tutorer les apprenants*<br>- *(C1010) Animer et piloter un parcours de formation de type CLOM/MOOC*<br>- *(C1089) Intégrer un contenu pédagogique au sein d'une plate-forme de formation en ligne et massive*<br>- *(C1117) Scénariser des séquences vidéos pédagogiques*<br>- *(C1043) Créer et animer une communauté de pratique* |
|---|---|

| *Définition* | Réalisation des mises à jour identifiées au cours de la mise en œuvre du dispositif de formation ; Suivi des travaux et activités des étudiants (analyse des traces d'activités) ; Tutorat et accompagnement des étudiants ; Création et animation d'une communauté d'apprenants. |
|---|---|
| *Mission associée* | - Mission C « Numérique éducatif » en collaboration avec la mission A « Enseignement supérieur et transfert d'expertises » |
| *Public cible* | - Acteurs éducatifs responsables (Doyens, chefs de départements, chefs d'entreprises, responsables CNF, etc.) |
| *Lieu(x) (accueil)* | - Phnom-Penh (regroupement régional des membres de l'équipe) |
| *Date(s)* | - Décembre 2017 |
| *Durée* | - 5 jours |
| *Type de formation* | - Présentiel |
| *Coût prévisionnel* | - Selon grille locale |
| *Indicateur* | - Pertinence des rapports |

Cette programmation pourrait être enrichie par d'autres types de formations (parcours existants ou ateliers à convenir à partir du référentiel AUF) en vue de consolider la dimension scientifique et de recherche, notamment dans son accompagnement aux formations doctorales dans les universités francophones qui le demanderaient.

Quelques exemples sont proposés ci-après :

## 2.3. Des activités pour l'appui à la recherche scientifique autour du numérique éducatif.

Les activités de l'appui à la recherche scientifique relèvent normalement de la mission B « recherche et thématiques intégrées », mais elles peuvent aussi être mutualisées avec les autres missions.

L'essentiel des activités pour l'appui à la recherche scientifique pour la période 2016-2017 serait composé de formation sur les compétences de la recherche scientifique : recherche/production/diffusion d'écris scientifiques et de produits de recherche.

Il existe déjà des thèmes dans l'ensemble des parcours de formation préparés par l'IFIC :

- Création de document scientifique avec LaTeX
- Piratage éthique
- Risques cybernétiques
- Veille informationnelle

Le référentiel des compétences de l'AUF pourvoit aussi un ensemble assez riche de thèmes de formation en faveur de la recherche scientifique. La proposition des formations suivantes correspondrait à l'état de la recherche observé dans les trois pays CLV et pourrait donc constituer, dans un esprit de continuum de l'activité de recherche/publication, une alternative de programmation dans la feuille de route du numérique éducatif pour 2016-2017.

**L. PARCOURS DE FORMATION : « RECHERCHE D'INFORMATION SCIENTIFIQUE »**

| | |
|---|---|
| *Titre(s) de compétence(s)* | - *(C1073) Formuler une question de recherche sur la base de sa connaissance de la littérature du domaine et de sa pratique du terrain*<br>- *(C1042) Consulter des contenus sous toutes les formes : texte, image, vidéo, etc.*<br>- *(C1130) Travailler, échanger, se divertir, rechercher l'information avec des applications nomades*<br>- *(C1061) Élaborer une stratégie de recherche d'information*<br>- *(C1090) Localiser l'information recherchée sur Internet*<br>- *(C1125) Synthétiser les résultats de recherche d'informations sur Internet*<br>- *(C1112) Rechercher l'information* |
| *Définition* | Cette formation multi-compétences vise à réaliser plusieurs activités (veille informationnelle et Recherche de l'information sur Internet) pour relever plusieurs défis scientifiques comme :<br><br>- La maîtrise des infrastructures et des organisations de la Web-science ;<br>- La maîtrise de toutes les dimensions de l'édition et de l'accès ouvert, aux documents scientifiques et pédagogiques : édition pour le Web, repérage des ressources ouvertes, accès et utilisation |
| *Mission associée* | - Mission B « Recherche et thématiques intégrées » en collaboration avec la mission A « Enseignement supérieur et transfert d'expertises » et C « Numérique éducatif » |
| *Public cible* | - Doctorants, ingénieurs de recherche, enseignants, chefs de laboratoire/équipes de recherche, responsables de publications scientifiques et profils assimilés. |
| *Lieu(x) (accueil)* | - Vientiane (regroupement régional) |
| *Date(s)* | - Février 2017 |
| *Durée* | - 5 jours |
| *Type de formation* | - Présentiel |
| *Coût prévisionnel* | - Selon grille locale |
| *Indicateur* | - Qualité des fils rouges (projets individuels) |

**M. PARCOURS DE FORMATION : « Rédaction scientifique »**

| | |
|---|---|
| *Titre(s) de compétence(s)* | - *(C1114) Rédiger des documents scientifiques* |
| *Définition* | Cette compétence permettrait de fournir des démarches et des outils pour appliquer les principes de la rédaction scientifique. Elle couvre plusieurs types de savoirs comme le fait de :<br><br>- Connaître les bases de l'édition multimédia (combinaison et spécificités des différents médias) |
|   | - Connaître les contraintes de publication online (chartes technique, graphique et éditoriale)<br>- Connaître les licences des droits d'auteurs et les code de la propriété intellectuelle<br>- Connaître les principes de la rédaction scientifique et les règles de l'écriture scientifique dans la structuration et la mise en forme<br>- Connaître les bases de l'éditique (édition papier et numérique, mise sous pli, diffusion) et la dématérialisation des documents : numérisation, capture LAD/RAD, workflow et intégration dans les Systèmes d'Information des utilisateurs<br>- Connaître les systèmes de gestion de références bibliographiques et les styles de rédaction de notices bibliographiques (Zotero, Endnote, etc.) |
| *Mission associée* | - Mission B « Recherche et thématiques intégrées » en collaboration avec la mission A « Enseignement supérieur et transfert d'expertises » et C « Numérique éducatif » |
| *Public cible* | - Doctorants, ingénieurs de recherche, enseignants, chefs de laboratoire/équipes de recherche, responsables de publications scientifiques et profils assimilés. |
| *Lieu(x) (accueil)* | - Danang (regroupement régional) |
| *Date(s)* | - Avril 2017 |
| *Durée* | - 5 jours |
| *Type de formation* | - Présentiel |
| *Coût prévisionnel* | - Selon grille locale |
| *Indicateur* | - Qualité des fils rouges (projets individuels) |

**N. PARCOURS DE FORMATION : « Communication des résultats de recherche »**

| *Titre(s) de compétence(s)* | - ***(C1023) Communiquer le résultat de ses recherches*** |
|---|---|
| *Définition* | Cette compétence couvre plusieurs types de savoirs comme le fait de :<br><br>- Connaître les organes de publication de sa communauté scientifique ainsi que les principales manifestations et colloques de celle-ci ;<br>- Connaître les règles de l'écriture scientifique ;<br>- Connaître les règles de l'expression et des présentations orales ;<br>- Connaître les règles de citation dans le texte ainsi que les principes de composition des références bibliographiques ;<br>- Connaître les nouvelles formes de publication en ligne, leurs formes juridiques correspondantes ;<br>- Connaître des logiciels de traitement de texte ;<br>- Connaître des logiciels de présentation ;<br>- Connaître des logiciels d'écriture et de travail collaboratif ;<br>- Connaître les obligations juridiques et les normes éthiques en matière de droits d'auteurs, de citation et de plagiat. |

| | | |
|---|---|---|
| *Mission associée* | - | Mission B « Recherche et thématiques intégrées » en collaboration avec la mission A « Enseignement supérieur et transfert d'expertises » et C « Numérique éducatif » |
| *Public cible* | - | Doctorants, ingénieurs de recherche, enseignants, chefs de laboratoire/équipes de recherche, responsables de publications scientifiques et profils assimilés. |
| *Lieu(x) (accueil)* | - | Ho-Chi-Minh (regroupement régional) |
| *Date(s)* | - | Novembre 2017 |
| *Durée* | - | 5 jours |
| *Type de formation* | - | Présentiel |
| *Coût prévisionnel* | - | Selon grille locale |
| *Indicateur* | - | Taux de satisfaction des participants |

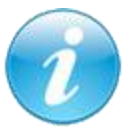

Toutes les formations de 2016-2017 sont destinées à former un potentiel stable en ressources humaines capable de reproduire la formation à l'échelle des pays. Il est donc indispensable d'impliquer des formateurs du pourtour professionnel et institutionnel des CNF et CNFp pour garantir la continuité des formations. Le but est de créer petit à petit des formateurs locaux parmi les enseignants et les futurs professionnels qui transmettraient les compétences acquises auprès des étudiants et dans les milieux professionnels, publics et privés.

| Lieu/mois | Hanoï | Danang | Ho-Chi-Minh | Phnom-penh | Vientiane |
|---|---|---|---|---|---|
| déc.-16 | Indexation des RELS | | Gouvernance d'un système d'information | | |
| janv.-17 | | Déploiement d'une plateforme de « eformation » Moodle | Conception de cours en ligne | Médiation d'un espace d'innovation | |
| févr.-17 | Etude et définition de son projet de dispositif de e-formation | | | Licence libre | Recherche d'information |
| mars-17 | Gestion de l'innovation | | Tutorat à distance | | |
| avr.-17 | | Rédaction scientifique | | FOAD Hybride | |
| mai-17 | Gérer la qualité informatique d'un SI | | | | Fondamentaux de la scénarisation pédagogique |
| juin-17 | | Administration d'une plate-forme FOAD | | Classe inversée | |
| sept.-17 | | | Animer et piloter un MOOC | | |
| oct.-17 | | | | | Évaluation dans les dispositifs d'apprentissage |
| nov.-17 | Intégration des ressources sur une PF de MOOC et gestion des fonctionnalités (Open EdX) | | Communication des résultats de recherche | | |
| déc.-17 | | | | Suivi pédagogique du dispositif de formation | |

Légende :
- Compétences professionnalisantes
- Compétences de diplomation
- Compétences de gouvernance
- Compétences de recherche

**Feuille de route proposée pour 2016-2017 : ateliers de formation sur les TIC/E**

## ACTIVITES

| | nov.-16 | déc.-16 | janv.-17 | févr.-17 | mars-17 | avr.-17 | mai-17 | juin-17 | sept.-17 | oct.-17 | nov.-17 | déc.-17 |
|---|---|---|---|---|---|---|---|---|---|---|---|---|
| « Accompagnement à l'usage des TIC/E » | | ■ | ■ | ■ | ■ | ■ | | | | | | |
| « Formations hybrides » | | ■ | ■ | ■ | ■ | ■ | | | | | | |
| « Création de cours en ligne structurés et interopérables » | | ■ | ■ | ■ | ■ | ■ | ■ | ■ | ■ | ■ | ■ | ■ |
| « RELs : Étude de l'existant » | ■ | ■ | ■ | | | | | | | | | |
| « RELs : collecte et traitement » | | | | ■ | ■ | ■ | ■ | ■ | ■ | | | |
| "RELs : diffusion et communication » ➔ | | | | | | | | | | ■ | ■ | ■ |

## ATELIERS DE FORMATION

| nov.-16 | déc.-16 | janv.-17 | févr.-17 | mars-17 | avr.-17 | mai-17 | juin-17 | sept.-17 | oct.-17 | nov.-17 | déc.-17 |
|---|---|---|---|---|---|---|---|---|---|---|---|
| | Indexation des RELS | Déploiement d'une plateforme de « eformation » Moodle | Etude et définition de son projet de dispositif de e-formation | Gestion de l'innovation | Rédaction scientifique | Gérer la qualité informatique d'un SI | Administration d'une plate-forme FOAD | Animer et piloter un MOOC | Évaluation dans les dispositifs d'apprentissage | Intégration des ressources sur une PF de MOOC et gestion des fonctionnalités (Open EdX) | Suivi pédagogique du dispositif de formation |
| | Gouvernance d'un système d'information | Médiation d'un espace d'innovation | Licence libre | Tutorat à distance | FOAD Hybride | Fondamentaux de la scénarisation pédagogique | Classe inversée | | | Communication des résultats de recherche | |
| | | Conception de cours en ligne | Recherche d'information | | | | | | | | |

**Feuille de route proposée pour 2016-2017 : répartition chronologique des activités et des formations**

## 3. POUR CONCLURE

Beaucoup de nouvelles décisions seront stratégiques pour l'AUF et ses bureaux régionaux en cette période de transition dans sa politique numérique. Nous tenons à rappeler que toute la région de l'Asie pacifique et notamment les pays CLV passe aussi par une phase de transition profonde dans le domaine des technologies éducatives, soutenue en cela par des alliances stratégiques régionales élaborées particulièrement avec l'ASEAN et ses structures et programmes éducatifs régionaux. Aussi, est-il fortement recommandé que la Francophonie prenne position par rapport à ces changements en cours au même titre que les autres nations comme la Corée du Sud, l'Australie, le Japon et les États-Unis.

Il y a certes une valorisation de l'action francophone déjà réalisée, mais cette action sera-t-elle suffisante pour constituer une alternative solide et compétitive qui pourrait rivaliser avec les actions d'autres acteurs comme l'ACU ? La Francophonie en Asie-Pacifique, par le témoignage de ses grands opérateurs, n'a pas vocation à jouer le même rôle qu'en Afrique ou en Europe de l'Est. Mais son patrimoine historique dans la région pacifique et ses acquis stratégiques valent beaucoup la peine d'investir davantage via le numérique éducatif pour entretenir une présence culturelle et linguistique stable et pérenne. Dans une région où une génération de francophones en positions de prise de décisions quitte progressivement la scène pour laisser place à une nouvelle génération de responsables moins attachés à la langue de Molière, est-elle source d'inquiétude ? Ce sont là des défis et des enjeux à regarder de plus près à travers le prisme de l'éducation et du numérique éducatif et qui deviennent de véritables enjeux régionaux. Ils constituent en effet un chapitre important dans la coopération universitaire qui se met actuellement en place dans le cadre de l'ASEAN. Si la Francophonie devrait garder un périmètre de présence académique et universitaire dans la région, elle devrait savoir composer avec les grands acteurs éducatifs de la région, notamment les structures de l'ASEAN comme KOIKA, l'AUN et l'ACU. Des voies de collaboration devraient être envisagées avec ces grands acteurs régionaux de l'éducation.